\documentclass[final,twoside,11pt]{article}
\usepackage{lscape}
\usepackage{ifthen,bm}
\usepackage{dsfont}
\usepackage{pgf}
\usepackage{xcolor}
\usepackage{latexsym, amsfonts, amssymb, amsmath}
\usepackage{alltt,latexsym, verbatim}
\usepackage{float}
\usepackage{epsfig}
\usepackage{graphicx}
\usepackage{amsfonts}
\usepackage{xspace}

\newtheorem{theorem}{Theorem}[section]
\newtheorem{lem}[theorem]{Lemma}
\newtheorem{pro}[theorem]{Proposition}
\newtheorem{cor}[theorem]{Corollary}
\newtheorem{conj}[theorem]{Conjecture}

\newtheorem{rem}[theorem]{Remark}
\newtheorem{com}[theorem]{Comments}
\newtheorem{ex}[theorem]{Example}
\newtheorem{defi}[theorem]{Definition}
\newtheorem{hyp}[theorem]{Assumption}

\newcommand{\bt}{\begin{theorem}}\newcommand{\et}{\end{theorem}}
\newcommand{\bl}{\begin{lem}}\newcommand{\el}{\end{lem}}
\newcommand{\bp}{\begin{pro}}\newcommand{\ep}{\end{pro}}
\newcommand{\bcor}{\begin{cor}}\newcommand{\ecor}{\end{cor}}
\newcommand{\bconj}{\begin{conj}}\newcommand{\econj}{\end{conj}}

\newcommand{\bd}{\begin{defi} \rm }\newcommand{\ed}{\end{defi} }
\newcommand{\brem }{\begin{rem} \rm }\newcommand{\erem }{\end{rem}}
\newcommand{\bcom}{\begin{com} \rm }\newcommand{\ecom }{\end{com}}
\newcommand{\brems }{\begin{rem} \rm }\newcommand{\erems }{\end{rem}}
\newcommand{\bex}{\begin{ex} \rm }\newcommand{\eex}{\end{ex}}
\newcommand{\bhyp}{\begin{hyp} \rm }\newcommand{\ehyp}{\end{hyp}}

\def \be{\begin{eqnarray}}
\def \ee{\end{eqnarray}}
\def \b*{\begin{eqnarray*}}
\def \e*{\end{eqnarray*}}

\def \LL{{\mathbb L}}
\def \MM{{\mathbb M}}

\def \H{{\bf H}}

\def \[{[\,\!\![}
\def \]{]\,\!\!]}

\def \1{{\bf 1}}

\def\F{{\cal T}}\def\F{{\cal G}}\def\F{{\cal F}}

\def\ttt{t \in [0,\Ts]}
\newcommand{\bea}{\begin{eqnarray*}}
\newcommand{\eea}{\end{eqnarray*}}
\newcommand{\beqa}{\begin{eqnarray}}
\newcommand{\eeqa}{\end{eqnarray}}

\def\w{\wedge}

\def\R{{\mathbb R}}

\def\I{\mathds{1}}
\def\sp{\,,\ \,}
\def\l{\label}
\def\r{\eqref}

\def\bal{\begin{aligned}}
\def\eal{\end{aligned}}

\def\ttt{{t \in [0,\Ts]}}
\def\ttd{{t \in [0,\tb]}}
\def\hat{\widehat}

\def\mym{m}

\def\xitau2{\xi_{(\thetau)}}\def\xitau2{\xi}
\def\xiitau2{\xi^i_{(\thetau)}}\def\xiitau2{\xi^i}
\def\xintau2{\tilde{\xi}_{(\thetau)}}\def\xintau2{\tilde{\xi}}
\def\chitau{\chi}

\def\Ltau2{\tilde{\xi}(\tau_{\mym},\tau_1)}
\def\chiitau2{P^i_{\tau}}

\newcommand{\Proba}{\mathbb{P}}

\newcommand{\lab }{\label }
\newcommand{\beq}{\begin{eqnarray*}}
\newcommand{\eeq}{\end{eqnarray*}}

\def\notilde{\tilde}\def\notilde{}
\def\mym{-}

\def\theb{b^+}\def\theb{b}
\def\thebm{b^-}\def\thebm{\bar{b}}
\def\thelambda{\lambda^+}\def\thelambda{\lambda}
\def\thelambdam{\lambda}\def\thelambdam{\lambda^-}\def\thelambdam{\bar{\lambda}}
\def\theg{g}

\def\mygb{\tilde{g}}
\def\mygh{\hat{g}}
\def\rt{\tilde{r}}
\def\betat{\tilde{\beta}}
\def\tilde{\widetilde}
\def\cadlag{c\`adl\`ag }

\def\unxi{\xi}
\def\thechit{\chi}

\def\lev{L\'{e}vy\xspace}
\def\ud{d}
\def\dt{\ud t}
\def\ds{\ud s}
\def\du{\ud u}
\def\dx{\ud x}

\def\dv{\ud v}

\def\gg{{\mathbb G}}\def\gg{{\mathbb F}}
\def\G{{\cal G}}\def\G{{\cal F}}
\def\ff{\mathbb{F}}
\def\Fs{\tilde{{\cal F}}}
\def\ff{\mathbb{F}^*}
\def\F{{\cal F}^*}

\def\gg{{\cal G}}\def\gg{\mathbb{G}}
\def\G{{\cal G}}
\def\ff{{\cal F}}\def\ff{\mathbb{F}}
\def\F{{\cal F}}
\def\hh{{\cal H}}\def\hh{\mathbb{H}}
\def\H{{\cal H}}

\def\Fs{\tilde{\cal F}}\def\Fs{{\cal E}}

\def\E{\mathbb{E}}
\def\Ef{\mathbb{F}}\def\Ef{\tilde{\mathbb{E}}}
\def\thebar{\bar}

\def\deltar{\epsilon}

\def\clc{\Rt^{i}}
\def\clf{\Rt^{f}}

\def\xipred{\bar{\xi}}\def\xipred{\tilde{\xi}}

\def\Rf{\mathfrak{R}}\def\Rf{\mathfrak{r}}

\def\hatp{p}
\def\hatq{\overline{\hatp}}

\def\Rfh{\hat{\Rf}}\def\Rfh{\Rf}

\def\tm{\tau_-}\def\tm{\overline{T}}
\def\tp{\tau_+}\def\tp{T}
\def\thetau{\vartheta}
\def\thetheta{\theta}

\def\theomega{\nu}\def\theomega{\tau}\def\theomega{\tau}

\def\Rt{\mathbf{R}}\def\Rt{\mathsf{R}}\def\Rt{R}

\def\tm{\tau_-}\def\tm{\overline{\theta}}
\def\tp{\tau_+}\def\tp{\theta}
\def\Ts{\bar{T}}\def\Ts{\mathcal{T}}\def\Ts{T}
\def\thetheta{\vartheta}
\def\tb{{\bar{\tau}}}
\def\theomega{\nu}\def\theomega{\tau}

\def\Ts{\mathcal{T}}\def\Ts{\mathsf{T}}\def\Ts{\bar{T}}\def\Ts{T}

\def\thisR{R}\def\thisR{\rho}
\def\thisRm{{\overline{\thisR}}}

\def\bl{\bea\bal}\def\el{\eal\eea}
\def\bll#1{\beqa\label{#1}\bal
}\def\lel{\eal\eeqa}

\newcommand{\ii}{\operatorname{i}\kern -0.8pt}

\def\P{\ensuremath{\mathrm{I\kern-.2em P}}}

\def\rc{}

\def\thealpha{\alpha}

\newcommand{\beql}[1]{\beqa\label{#1}\begin{aligned}}
\newcommand{\eeql}{\eal\eeqa}
\newcommand{\bel}{\bea\bal}
\newcommand{\eel}{\end{aligned}\eea}

\title{{\Large \bf
Counterparty Risk and Funding:\\
The Four Wings of the TVA
}}

\author{St\'{e}phane Cr\'{e}pey, R\'{e}mi Gerboud, Zorana Grbac and Nathalie Ngor \footnote{The research of S. Cr\'{e}pey and Z. Grbac benefited from the support of the `Chaire Risque de cr\'edit',
F\'ed\'eration Bancaire Fran\c caise. R. Gerboud and N. Ngor are
former students from the MSc quantitative finance program {M2IF} at Evry University. The authors warmly
thank Anthony Dorme and Remi Takase, two other former M2IF students, for their help in the numerical studies of the \lev model and of the TVA BSDEs.}
\\\\
Laboratoire Analyse et Probabilit\'es \\ Universit\'e d'\'Evry Val d'Essonne \\
91037 \'Evry Cedex, France}

\date{\today}

\begin{document}
\thispagestyle{empty}
\maketitle
\thispagestyle{empty}
\begin{abstract}
The credit crisis and the ongoing European sovereign debt crisis have highlighted the native form of credit risk,
namely the counterparty risk. The related {Credit Valuation Adjustment
 (CVA),
 Debt Valuation Adjustment
 (DVA),
Liquidity} Valuation Adjustment
 {(LVA)} and Replacement Cost (RC) issues, jointly  referred to in this paper as Total Valuation Adjustment (TVA), have been thoroughly investigated in {the} theoretical papers
\cite{Crepey11b1} and \cite{Crepey11b2}. The present work provides an executive summary and numerical companion to these papers,
through which the TVA pricing problem can be reduced to Markovian pre-default {TVA} BSDEs.
The first step consists in the counterparty
clean valuation {of a portfolio of contracts}, which is {the} valuation in {a} hypothetical situation where the two parties would be risk-free and funded at a risk-free rate.
In {the} second step, the TVA is obtained as the value of an option on the
counterparty clean value process {called Contingent Credit Default Swap (CCDS)}.
{Numerical results are {presented} for interest rate swaps in the Vasicek, as well as in the inverse Gaussian Hull-White short rate model, also allowing one to assess the related model risk issue.}
\end{abstract}
\begin{keywords}
Counterparty risk, Credit valuation adjustment (CVA),
Debt valuation adjustment (DVA), Liquidity valuation adjustment (LVA),
Replacement cost (RC),
Backward stochastic differential equation (BSDE), \lev process, Interest rate swap.
\end{keywords}

{\def\lk{L\'{e}vy--Khintchine\xspace}
\newcommand{\prozess}[1][L]{{\ensuremath{#1=(#1_t)_{0\le t\le T_{*}}}}\xspace}
\newcommand{\prazess}[1][L]{{\ensuremath{#1=(#1_t)_{t\ge0}}}\xspace}

\newcommand{\la}{\langle}
\newcommand{\ra}{\rangle}

\def\EM{\ensuremath{(\mathbb{EM})}\xspace}
\def\SUP{\ensuremath{(\mathbb{SUP})}\xspace}
\def\B{\ensuremath{(\mathbb{B})}\xspace}
\def\Bzero{\ensuremath{(\mathbb{B'})}\xspace}
\def\Hone{\ensuremath{(\mathcal{H}1)}\xspace}
\def\Htwo{\ensuremath{(\mathcal{H}2)}\xspace}
\def\Hthree{\ensuremath{(\mathcal{H}3)}\xspace}
\def\Hfour{\ensuremath{(\mathcal{H}4)}\xspace}
\def\lone{\ensuremath{(\mathbb{L}.1)}\xspace}
\def\ltwo{\ensuremath{(\mathbb{L}.2)}\xspace}
\def\rlone{\ensuremath{(\mathbb{RL}.1)}\xspace}
\def\rltwo{\ensuremath{(\mathbb{RL}.2)}\xspace}

\def\MM{\ensuremath{\mathscr{M}}}
\def\ME{\mathbb{M}}

\def\bD{\mathbb{D}}

\def\R{\ensuremath{\mathbb{R}}}
\def\Rp{\mathbb{R}_{\geqslant0}}
\def\Rm{\mathbb{R}_{\leqslant 0}}
\def\C{\mathbb{C}}
\def\U{\ensuremath{\mathcal{U}}}

\def\e{\mathrm{e}}

\def\P{\ensuremath{\mathrm{I\kern-.2em P}}}
\def\Q{\mathbb{Q}}
\def\E{\ensuremath{\mathrm{I\kern-.2em E}}}

\def\e{\mathrm{e}}

\newcommand{\set}[1]{\ensuremath{\left\{#1\right\}}}

\newcommand{\cB}{{\mathcal{B}}}
\newcommand{\cC}{{\mathcal{C}}}
\newcommand{\cE}{{\mathcal{E}}}
\newcommand{\cD}{{\mathcal{D}}}
\newcommand{\cM}{{\mathcal{M}}}
\newcommand{\cO}{{\mathcal{O}}}
\newcommand{\cP}{{\mathcal{P}}}
\newcommand{\ha}{{\mathbb{H}}}

\newcommand{\indik}{{\mathbf{1}}}
\newcommand{\ifdefault}[1]{\ensuremath{\mathbf{1}_{\{\theomega^{*} \leq #1\}}}}
\newcommand{\ifnodefault}[1]{\ensuremath{\mathbf{1}_{\{\theomega^{*} > #1\}}}}

\def\notilde{}

\def\EM{\ensuremath{(\mathbb{EM})}\xspace}
\def\ES{\ensuremath{(\mathbb{ES})}\xspace}
\def\AC{\ensuremath{(\mathbb{AC})}\xspace}
\def\LL{\ensuremath{(\mathbb{L})}\xspace}
\def\VOL{\ensuremath{(\mathbb{VOL})}\xspace}

\def\utop{u^{\top}}\def\utop{u}
\def\ztop{z}

\def\thenu{\tau}
\def\mygp{g}

\def\cure{\eta}\def\cure{c}\def\cure{\bar{\delta}}\def\cure{\epsilon}
\def\thecure{\delta}\def\thecure{\delta}

\newcommand{\srem}[1]{\marginpar{{\scriptsize  #1 }}}

\section{Introduction}

The credit crisis and the ongoing European sovereign debt crisis have highlighted the native form of credit risk, namely the counterparty risk. This is the risk of non-payment of promised cash-flows due to the default of the counterparty in a bilateral OTC derivative transaction. The basic counterparty risk mitigation tool is a Credit Support Annex (CSA) specifying a valuation scheme of {a portfolio of contracts at the default time of a party.  In particular, the netting rules which will be applied to the portfolio are specified, as well as a collateralization scheme similar to margin accounts in futures contracts.}

By extension counterparty risk is also the volatility of the price of this risk, {this price being known as} Credit Valuation Adjustment (CVA), see \cite{DePriscoRosen05}, \cite{BrigoMoriniPallavicini12}, \cite{BieleckiBrigoCrepeyHerbertsson13}. {Moreover, as banks themselves have become risky, counterparty risk must be understood in a bilateral perspective, where the counterparty risk of the two parties are jointly accounted for in the modeling. Thus, in addition to the CVA, a Debt Valuation Adjustment (DVA) must be considered. }In this context the classical assumption of a locally risk-free asset which is used for {financing purposes of the bank} is not sustainable anymore. This raises the companion issue {of proper} {nonlinear} accounting of the funding costs of a position, and a corresponding Liquidity Valuation Adjustment (LVA). {The} last related issue is that of Replacement Cost (RC) corresponding to the fact that at the default time of a party the contract is valued by the liquidator according to a CSA valuation scheme which can fail to reflect the actual value of the contract at that time.

The CVA/DVA/LVA/RC intricacy issues, jointly referred to henceforth as TVA (for Total Valuation Adjustment), were thoroughly investigated in the theoretical papers
\cite{Crepey11b1}, \cite{Crepey11b2}, \cite{BrigoCapponiPallaviciniPapatheodorou11} and \cite{BurgardKjaer10}. The present work is a numerical companion paper to \cite{Crepey11b1} and \cite{Crepey11b2}.
Section \ref{s:cva} provides
an executive summary of the two papers.
Section \ref{s:csa} describes {various} CSA specifications.
In Sections \ref{s:clean} and \ref{s:cvacomp} we present clean valuation (clean of counterparty risk and funding costs) and TVA computations in {two simple models} for interest rate derivatives. {We show {in a series of} practical examples
how CVA, DVA, LVA and RC can be computed in various situations,
and we also assess the related {model risk} issue.}

\section{TVA Representations}\label{s:cva}

\subsection{Setup}
We consider a netted portfolio of OTC derivatives between two defaultable counterparties, generically referred to
as the ``contract between the bank and her counterparty''. {The counterparty is most commonly a bank as well.}
Counterparty risk and funding cash-flows can only be considered at a netted and global level, outside
the scope of different business desks. Consequently, {the price $\Pi$ of the contract must be computed as {a} difference between the clean price $P$ provided by the relevant business desk  and a correction $\Theta$ computed by {the} central TVA desk.
By {price} we mean here the cost for the bank of margining, hedging and funding
(``cost of hedging'' for short). By clean price we mean {the price of the contract computed without taking into account counterparty risk and excess funding costs.}
{Symmetrical} considerations apply to the counterparty, but with non-symmetrical data in the sense of hedging positions and funding conditions.
As a consequence and due to nonlinearities in the funding costs, the prices (costs of
hedging) are not the same {for} the two parties.
For clarity we focus on {the bank's
price} in the sequel.

We denote by $\Ts$ the time horizon of the  contract with promised dividends
$dD_t$ from the bank to her counterparty. {Both} parties are defaultable, with respective {default times}
denoted by $\tp$ and $\tm$. This results in an effective dividend stream
$dC_t=J_t dD_t$, where $J_t=\I_{t<\tau}$ with $\theomega=\tp\w\tm$.
One denotes by $\tb=\theomega\wedge \Ts$
the effective time horizon of {the contract} as there are no cash-flows after $\tb$.
The case of unilateral counterparty risk (from the perspective of the bank) can be recovered by letting $\tp\equiv \infty$. After having sold
the contract to the investor at time 0,
the bank sets-up a
collateralization, hedging and funding
portfolio (``hedging portfolio'' for short).
We call an external funder of the bank (or funder for short) a generic third-party, possibly composed in practice of several entities or devices,
insuring funding of the position of the bank. This funder, assumed default-free for simplicity, thus plays the role of ``lender/borrower of last resort'' after exhaustion
 of the {internal} sources of funding provided to the bank via the dividend and funding gains on her hedge, or via the remuneration of the margin amount.

The full model filtration is given as
$\gg = \ff \vee \hh^{\tp} \vee {\hh}^{\tm}$,
where $\ff$ is a reference filtration and $\H^{\tp}_t=\sigma(\tp\wedge t),$
${\H}^{\tm}_t=\sigma(\tm\wedge t)$.
A probability space $(\Omega,\G_T,\mathbb{P})$,
{where $\mathbb{P}$ is some risk-neutral pricing measure},
is fixed throughout. The meaning of a risk-neutral pricing measure
in this context, with different funding rates in particular (see \cite{Crepey11b1}),
will be specified by martingale conditions introduced below in the form
of suitable pricing backward stochastic differential equations {(BSDEs); see
\cite{ElkarouiPengQuenez97} for {the} seminal reference in finance.}
\brem
Even though it will not appear explicitly in this paper, a pricing measure must also be such that the gain processes on the hedging assets follow
martingales, see \cite{Crepey11b1}.
\erem
Moreover, we assume that $\ff$ is immersed into $\gg$ in the sense that
an ${\ff}$-martingale
stopped at ${\theomega}$ is a $\gg$-martingale.
\brem As discussed in \cite{Crepey11b2},
this basic assumption precludes major wrong-way risk effects such as the ones
which occur with counterparty risk on credit derivatives. In particular, under these assumptions
an ${\ff}$-adapted \cadlag process cannot jump at $\tau.$ {We refer to \cite{Crepey13} for an extension of}
the general methodology of this paper beyond the {above immersion setup to incorporate wrong-way risk when necessary.}
\erem

\subsection{Data}

We denote by $r_t$ an
OIS rate, {where OIS stands {for an} Overnight Indexed Swap}, the best market proxy of a risk-free rate.
{By $\rt_t=r_t+ \gamma_t$  we denote} the credit-risk adjusted rate, where
$\gamma_t$ is the {$\ff$-{hazard intensity}
of $\theomega$, {which is assumed to exist.}
Let $\beta_t=\exp(-\int_0^t  r_s  ds)$ and
$\betat_t=\exp(-\int_0^t  \rt_s  ds)$
stand for the corresponding discount factors.
{Furthermore, $\E_{t}$
and $\Ef_{t}$ stand for the conditional expectations given
$\G_{t}$ and ${\F}_{t},$ respectively.}
The clean value process $P_t$ of the contract with promised dividends $dD_t$
is defined as, for $\ttd$,
\beqa\label{cleanf}{\beta} _t P_t:=
{\Ef}_t \left( \int_t^{\Ts} {\beta} _s dD_s \right)=
{\E}_t \big[\int_t^\tb  {\beta}_s dD_s + \beta_\tb P_\tb \big]\eeqa
by immersion of $\ff$ into $\gg$.

Effectively, promised dividends $dD_t$ stop being paid at $\tau$
(if $\tau<T$),
at which time the last terminal cash-flow $\Rt$
paid by the bank  closes out her position.
Let
an $\ff$-adapted process $Q$ represent the (predictable) CSA value process of the contract,
and an $\ff$-adapted process $\Gamma$ stand for the value process of a CSA (cash) collateralization scheme.
We denote by $\pi$ a real number meant to represent the wealth of the hedging portfolio of the bank
in the financial interpretation.
The close-out cash-flow $\Rt =\Rt(\pi)$ is in fact twofold, decomposing into
a close-out cash-flow $\clc$ from the bank to the counterparty, minus, in case of default
of the bank, a cash-flow $\clf=\clf(\pi)$ from the funder to the bank (depending on $\pi$).
These two
cash-flows are respectively derived from
the algebraic debt $\chi$ of the bank
to her counterparty  and $\mathfrak{X}(\pi)$ of the bank to her funder,
  modeled at time $\tau$ as \beqa\label{dette}\chi=Q_{\tau} - \Gamma_\tau\sp \mathfrak{X}(\pi)=-(\pi  - \Gamma_{\tau-} ). \eeqa
The {close-out cash-flow},
if $\tau<T$, is then modeled as $\Rt (\pi)=\clc - \I_{\tau=\tp}\clf (\pi)$, where
\begin{eqnarray}\label{e:divsup}\begin{aligned}
&\clc = \Gamma_\tau     +
\I_{\tau=\tp}\big(
\thisR  \chitau^{+}- \chitau^{-}\big)
-\I_{\tau=\tm}\big(
\thisRm \chitau^{-}- \chitau^{+}\big)
- \I_{\tp=\tm}\chitau\\
&\clf (\pi)=(1-\Rf)\mathfrak{X}^{+}(\pi),
\end{aligned}
\end{eqnarray}
in which $\thisR$ and $\thisRm$ stand for recovery rates
between the two parties, and $\Rf$ stands for a recovery rate of the bank to her funder.
The $\G_{\tau}$-measurable exposure at default is defined in terms of $R$ as
\beqa\label{d:unxi}\unxi (\pi):=
P_\tau
-\Rt(\pi)
=P_{\tau} - Q_{\tau} +   \I_{\tau= \tp} \big((1-\thisR)\chi^{+} +(1-\Rf)\mathfrak{X}^{+}(\pi) \big)-
(1-\thisRm)   \I_{\tau= \tm } \chi^{-},\eeqa
{where the second equality follows by an easy algebraic manipulation.}

We now consider the cash-flows required for funding the bank's position, meant in the sense of the contract and its hedging portfolio altogether. For simplicity we stick to the most common situation where the hedge is self-funded as swapped and/or traded via repo markets (see \cite{Crepey11b1}).
The OIS rate
$r_t$ is used as a reference for all other funding rates, which are thus defined in terms of corresponding bases to $r_t.$
Given such bases $b_t$ and $\bar{b}_t$ related to the collateral posted and received by the bank,
and $\thelambda_t$ and $\thelambdam_t$ related to external lending and borrowing,
the funding coefficient
$\theg_t(\pi)$
is defined by
\beqa\label{eqfs}\bal
&\theg_t(\pi)=  (b_t \Gamma_t^{+} -  \bar{b}_t \Gamma_t^{-})
+ {\thelambda}_t  \left(\pi- \Gamma_t \right)^{+}
-{\thelambdam}_t  \left(\pi- \Gamma_t \right)^{-}.
\eal\eeqa
Then $(r_t \pi+g_t(\pi))dt$
represents the bank's funding cost over $(t,t+dt),$
depending on its wealth $\pi$.
\brem
A funding basis is typically interpreted as a combination of liquidity and/or credit risk, see
\cite{FilipovicTrolle11} {and} \cite{CrepeyDouady12}. Collateral posted in foreign-currency and switching currency collateral optionalities can also be accounted for by suitable amendments to $b$ and $\bar{b},$
see \cite{FujiiShimadaTakahashi10b} {and} \cite{Piterbarg12}.
\erem

\subsection{BSDEs}

With the data
$\xi$ and $g$  specified,
the TVA process ${\Theta}$ can be {implicitly defined on $[0,\tb]$
as the solution to the following BSDE, posed in integral form and over the random time interval
$[0,\tb]$:}
For $\ttd$,
\beqa\lab{nice}
{\beta}_t {\Theta}_t= \E_t\Big[{\beta}_{\tb}\I_{\theomega <\Ts}\xi(P_\tau -\Theta_{\tau-} ) +\int_t^{\tb}{\beta}_{s} \theg_s (P_s -\Theta_s)
ds \Big].
\eeqa
The reader is referred to \cite[Proposition 2.1]{Crepey11b2}
for the derivation of the TVA {BSDE}
\eqref{nice}. In this paper for simplicity of presentation we take
\eqref{nice} as the
definition of the TVA.

The practical conclusion of \cite{Crepey11b2} is that
one can even adopt a simpler ``reduced, pre-default'' perspective
in which defaultability of the two parties only shows up through
their default intensities, {see Proposition
\ref{p:pred} below  and equation (3.8) in \cite{Crepey11b2}}.
For $\ttt$ and $\pi\in\R$ let
\beqa\label{eqgrebis}\bal
\thechit_t &= Q_{t}  -     \Gamma_t \\
\tilde{\xi}_t(\pi) &
 =   \left( P_{t} - Q_{t}\right) + \hatp_t  \big((1-\thisR_t) \thechit_t^{+}
 +(1-\Rfh )
(\pi  -     \Gamma_t)^{-}(\pi)
 \big)-
\hatq_t (1-\thisRm_t)   \thechit_t^{-},
\eal
\eeqa
in which
\bl
&\hatp_{\theomega} =\Proba(\theomega=\tp \,|\,\mathcal{G}_{\theomega-})\sp\hatq_{\theomega} =\Proba(\theomega=\tm \,|\,\mathcal{G}_{\theomega-}).\el
Note that in case of unilateral counterparty risk,
we have $\tp=\infty $ and consequently, $\hatp_{\theomega} =0$, $\hatq_{\theomega} =1.$

\begin{pro}[TVA Reduced-Form Representation]
\label{p:pred}
One has $\Theta=\tilde{\Theta}$ on $[0,\tb)$
and ${\Theta}_\tb= \I_{\theomega<\Ts} \unxi$,
where
 \beqa\lab{nicebis}
{\betat}_t \tilde{{\Theta}}_t= \Ef_t   \Big[ \int_t^{\Ts}{\betat}_{s}  \big(g_s(P_s - \tilde{\Theta}_s)
+\gamma_s \tilde{\xi}_s(P_s - \tilde{\Theta}_s) \big)ds \Big],
\eeqa
{for $\ttt$.}
\ep

\brem
This assumes that the data of \eqref{nicebis} are in $\ff,$ a mild condition which can always be met be passing to $\ff$-representatives (or pre-default values) of the original data.
\erem

\begin{rem}\rm
\label{rem:lincase}
For
$\Rf=1$ the exposure at default $\xi$ does not depend on $\pi,$
and in case of a linear funding coefficient given as
$g_t(P-\thetheta)=g^{0}_t(P) - \lambda^{0}_t \thetheta,$ {for some $g^{0}$ and $\lambda^{0}$},
the TVA equations \r{nice}
and
\r{nicebis}
respectively boil down to
the explicit representations
\beqa\l{e:first}
&{\beta}^{0}_t {\Theta}_t= \E_t \Big[\beta^{0}_{\tb} \I_{\theomega <\Ts}\xi +\int_t^{\tb}{\beta}^{0}_{s}  \theg_s^{0} (P_s)
ds  \Big] \\\l{e:second}
& \tilde{\beta}^{0}_t \tilde{{\Theta}}_t= \tilde{\E}_t \Big[ \int_t^{\Ts}\tilde{{\beta}}^{0}_{s} \Big(\theg^{0}_s(P^{0}_s)
+\gamma_s \xipred_s^{0}\Big)ds  \Big].
\eeqa
Here the funding-adjusted discount factors are
\bea\label{e:notcred+}
\bal
&{\beta}^{0}_t=\exp(-\int_0^t (r_s + \lambda^{0}_s) ds)\sp \tilde{{\beta}}^{0}_t=\exp(-\int_0^t (\rt_s + \lambda^{0}_s) ds)
\eal
\eea
and $\xipred_t^{0}$ in \r{e:second} is given by
$$\xipred_t^{0}
 =  \left( P_{t} - Q_{t}\right) + \hatp_t  (1-\thisR_t) \thechit_t^{+}
 -
\hatq_t (1-\thisRm_t)  \thechit_t^{-}.$$
On the numerical side
such explicit representations allow one to estimate the corresponding ``linear TVAs''
by standard Monte Carlo loops provided that $P_t$ and $Q_t$ can be computed explicitly. In general (for example as soon as ${\mathfrak{r}}<1$), nonlinear TVA computations
%{in Markov setups}
can only be done by more advanced schemes involving
linearization \cite{FujiiTakahashi11},  nonlinear regression \cite{Cesarietal10} or branching particles \cite{HenryLabordere12}.
 Deterministic schemes for the corresponding semilinear TVA PDEs can only be used in
low dimension.
\erem

In differential form the pre-default TVA BSDE \eqref{nicebis} {reads} as follows (cf. \cite[Definition 3.1]{Crepey11b2})
\begin{equation}\label{e:predcvabsddiscintrinsic}\begin{array}{c}\left\{\begin{aligned}
&\tilde{\Theta}_{\Ts}=0, \mbox{and for }\ttt:\\
&-d\tilde{\Theta}_t = \mygb_t(P_t -\tilde{\Theta}_t)dt-  d\tilde{\mu}_t ,
 \end{aligned}\right.\end{array}\end{equation}
 where $\tilde{\mu}$ is the $\ff$-martingale component of $\tilde{\Theta}$
 and
\beqa\lab{fdivbila}\bal
 \mygb_t (P_t -\vartheta)
  &=g_t(P_t - \vartheta)
+  \gamma_t \tilde{\xi}_t(P_t - \vartheta) -  \rt_t   \vartheta.
\eal
\eeqa

\brem
From \eqref{cleanf},  $P$ satisfies the following $\ff$-BSDE
\begin{equation}\label{e:bsdeclean}\begin{array}{c}\left\{\begin{aligned}
&P_{\Ts}=0, \mbox{and for }\ttt:\\
&-dP_t = dD_t - r_tP_t dt-  dM_t,
 \end{aligned}\right.\end{array}\end{equation}
for some $\ff$-martingale $M$. {Therefore,} the following pre-default
$\ff$-BSDE in $\tilde{\Pi}:=P-\tilde{\Theta}$
follows from \eqref{e:predcvabsddiscintrinsic} and
\eqref{e:bsdeclean}:
\begin{equation}\label{e:pi}\begin{array}{c}\left\{\begin{aligned}
&\tilde{\Pi}_{\Ts}=0, \mbox{and for }\ttt:\\
&-d\tilde{\Pi}_t = dD_t - \left(\mygb_t(\tilde{\Pi}_t)+r_tP_t\right)dt- d\tilde{\nu}_t,
 \end{aligned}\right.\end{array}\end{equation}
where $d\tilde{\nu}_t =dM_t - d\tilde{\mu}_t .$ As the pre-default price BSDE \eqref{e:pi} involves
the contractual promised cash-flows $dD_t$, it is less user-friendly than
the pre-default TVA BSDE \eqref{e:predcvabsddiscintrinsic}. This mathematical incentive
comes on top of the financial justification recalled at
the beginning of Section \ref{s:cva}
for adopting a two-stage ``clean price $P$ minus TVA correction $\Theta$'' approach
to the counterparty risk and funding issues.
 \erem

\subsubsection{Pre-default Markov Setup}
\label{ss:markov}

Assume
\beqa\l{eq:mark}\mygb_t(P_t-\vartheta)=
\mygh(t,X_t,\theta)\eeqa
for some deterministic
function
$\mygh(t,x,\theta),$
and an
$\R^d$-valued
$\ff$-Markov pre-default factor process $X$.
Then $\tilde{\Theta}_t=\tilde{\Theta}(t,X _t),$
where the {pre-default {TVA} pricing function}
$\tilde{\Theta}(t,x)$
is the solution to the following
{pre-default pricing PDE}:
\begin{equation}\label{theVI}\begin{array}{c}\left\{\begin{aligned}
&\tilde{\Theta}(\Ts,x)=0\,,\; x \in \mathbb{R}^{d} \\
&\left(\partial_t + {\mathcal{X}}\right) \tilde{\Theta} (t,x) +\mygh (t,x,\tilde{\Theta}(t,x))
 = 0 \mbox{ on } [0,\Ts)\times \mathbb{R}^{d},
 \end{aligned}\right.\end{array}\end{equation}
in which $\mathcal{X}$ stands for
the infinitesimal generator {of $X$}.
{As mentioned in Remark \ref{rem:lincase}}, from the point of view of numerical solution, deterministic PDE schemes for \eqref{theVI} can be used provided
the dimension of $X$ is less than 3 or 4, otherwise simulation schemes for \eqref{e:predcvabsddiscintrinsic}
 are the only viable alternative.

\subsection{CVA, DVA, LVA and RC}
\label{ss:wings}
Plugging \r{eqgrebis} into \r{fdivbila}
and reordering terms {yields}
%that
\bll{fdivbila1}
\mygb_t (P_t-\vartheta)+r_t \vartheta       = &-
\gamma_t \hatq_t(1-\thisRm)
(Q_{t}
-     \Gamma_t )^{-}
\\&  +\,
\gamma_t \hatp_t
\big((1-\thisR)(Q_{t}
-     \Gamma_t )^{+}
\\& \, +\,
 \theb_t \Gamma_t^{+}-
 \thebm _t \Gamma_t^{-} \,+\,  \thelambda_t (P_t -\vartheta- \Gamma_t )^{+}   \,-\,
\tilde{\lambda}_t  (P_t -\vartheta- \Gamma_t )^{-}  \\
&  +\,\gamma_t
\left( P_{t} -  \vartheta - Q_t \right),
\lel
where the coefficient $\tilde{\lambda}_t:={\thelambdam}_t -\gamma_t \hatp_t (1-\Rf) $ of $\left(P_t -\vartheta- \Gamma_t\right)^{-}$ in the third line
can be interpreted as an external borrowing basis net of credit spread. This coefficient represents the liquidity component of $\thelambdam$.
From the perspective of the bank,
the four terms
in this decomposition
 of the TVA $\Theta$
 can respectively be interpreted as a costly
 (non-algebraic, strict)
 credit value adjustment (CVA), a beneficial debt value adjustment (DVA),
 a liquidity funding benefit/cost (LVA), and a replacement benefit/cost (RC).
 In particular, the time-0 TVA can be represented as
 \beql{nicev}
{{\Theta}}_0
= &-\E \Big[\int_0^{\Ts}
{\beta}_{t}\gamma_t (1-\thisRm)\hatq_t
(Q_{t}
-     \Gamma_t )^{-}dt \Big]
\\&  +\,\E \Big[ \int_0^{\Ts}\,
{\beta}_{t}\gamma_t (1-\thisR)\hatp_t
 (Q_{t}
-     \Gamma_t )^{+}dt \Big]
\\&  +\,\E \Big[ \int_0^{\Ts}{\beta}_{t}
\big(\theb_t \Gamma_t^{+}-
 \thebm _t \Gamma_t^{-} \,+\,  \thelambda_t (P_t -\tilde{\Theta}_t- \Gamma_t )^{+}   \,-\,
\tilde{\lambda}_t  (P_t -\tilde{\Theta}_t- \Gamma_t )^{-} \big) dt  \Big] \\
&  +\,\E \Big[ \int_0^{\Ts} {\beta}_{t}\gamma_t
 ( P_{t} -  \tilde{\Theta}_t - Q_t  )dt \Big].
\eeql
The DVA and the $\gamma_t \hatp_t (1-\Rf) (P_t -\tilde{\Theta}_t- \Gamma_t )^{-}$-component of
the LVA can be considered as ``deal facilitating'' as  they increase the TVA and therefore decrease the price (cost of the hedge) the bank can consider selling the contract to her counterparty.
Conversely, the CVA and the $\thelambdam_t (P_t -\tilde{\Theta}_t- \Gamma_t )^{-}$ components of the LVA (for $\thelambdam_t$ positive) can be considered as ``deal hindering'' as  they decrease the TVA and therefore increase the price (cost of the hedge) for the bank.
The remaining terms can be interpreted likewise as ``deal facilitating or hindering'' depending on their sign, which is unspecified in general.

\section{CSA Specifications}\label{s:csa}

In the next subsections we detail various specifications
of the general form \eqref{fdivbila1} of $\mygb$, depending
on the CSA data: the close-out valuation scheme $Q,$ the collateralization scheme $\Gamma$ and the collateral remuneration bases $b$ and $\bar{b}$.

\subsection{Clean CSA Recovery Scheme}\label{ss:cleanCSA}
 In case of a clean CSA recovery scheme $Q=P$,
\r{fdivbila1} rewrites as follows:
\bll{fdivbila2}
\mygb_t (P_t-\vartheta)+\rt_t \vartheta        =& -\,
\gamma_t \hatq_t(1-\thisRm)
(P_{t}
-     \Gamma_t )^{-}
\\&+\,\gamma_t \hatp_t
\big((1-\thisR)(P_{t}
-     \Gamma_t )^{+}  \\
&  +\,
 \theb_t \Gamma_t^{+}-
 \thebm _t \Gamma_t^{-} \,+\,  \thelambda_t (P_t -\vartheta- \Gamma_t )^{+}   \,-\,
\tilde{\lambda}_t  (P_t -\vartheta- \Gamma_t )^{-}  .
\lel
Note $\tilde{r}_t$ on the left-hand side as opposed to ${r}_t$  in \eqref{fdivbila1}.
In case of
no collateralization, i.e. for $\Gamma=0$, the right-hand-side of \eqref{fdivbila2}
reduces to
\bll{fdivbila3}
  -
\gamma_t \hatq_t(1-\thisRm)
P_{t}^{-} \,+\,\gamma_t \hatp_t
(1-\thisR) P_{t}^{+}  +\,
 \thelambda_t \left(P_t -\vartheta\right)^{+}
\, -\,
\tilde{\lambda}_t \left(P_t  -\vartheta\right)^{-};
\lel
whereas in case of continuous collateralization with $\Gamma=Q=P$,
it boils down
to
 \bll{fdivbila4}
 -
 \thebm _t P_t^{-} + \theb_t P_t^{+} +  \thelambda_t \vartheta^{-}
-
\tilde{\lambda}_t \vartheta^{+}.
\lel

\begin{rem}\rm
\label{rem:lincase-red-clean}
If $\thelambda =\tilde{\lambda}$ (case of equal external borrowing and lending liquidity bases),
the TVA is linear
for every collateralization scheme of the form
$P_{t} -     \Gamma_t =\varepsilon_t$,
for some exogenous\footnote{Not depending on $\tilde{\Theta}_t$, like with null or continuous collateralization.}
residual exposure $\varepsilon_t$, {cf. Remark \ref{rem:lincase}.
Setting}
\bll{fdivbila2bis}
\mygb^{\lambda}_t    &=
-\left(\gamma_t \hatq_t(1-\thisRm) +{\lambda}_t \right)\varepsilon^-_t
+
\left(\gamma_t \hatp_t
(1-\thisR)+\lambda_t \right)\varepsilon^+_t
+ \theb_t \Gamma_t^{+}
\, -\,  \thebm _t \Gamma_t^{-} ,
\lel
one ends up, {similarly to \eqref{e:second},} with
\beqa\l{e:firstbis}
& \tilde{\beta}^{\lambda}_t \tilde{{\Theta}}_t= \tilde{\E}_t   \int_t^{\Ts}\tilde{{\beta}}^{\lambda}_{s} \mygb^{\lambda}_s   ds
\eeqa
for the funding-adjusted discount factor
\beqa\label{e:notcredquattro}
\bal
&\tilde{{\beta}}^{\lambda}_t=\exp(-\int_0^t (\rt_s + \lambda_s) ds).
\eal
\eeqa
\erem

\subsection{Pre-Default CSA Recovery Scheme}\label{ss:predCSA}

In case of a pre-default CSA recovery scheme $Q=\tilde{\Pi}=P-\tilde{\Theta}$,
\r{fdivbila1} rewrites as follows
\bll{fdivbila5}
\mygb_t (\vartheta)+r_t \vartheta        =& -
\left(\gamma_t \hatq_t(1-\thisRm)+\tilde{\lambda}_t\right)
(P_{t}-\vartheta
-     \Gamma_t )^{-}
\\&+ \left(\gamma_t \hatp_t
(1-\thisR)+\thelambda_t\right)(P_{t}-\vartheta
-     \Gamma_t )^{+}  \\
&+
 \theb_t \Gamma_t^{+}
-
 \thebm _t \Gamma_t^{-}.
\lel
In case of
no collateralization, i.e. for $\Gamma=0$, the right-hand-side
reduces to
\bll{fdivbila6}
-\,
 \big(
\gamma_t \hatq_t(1-\thisRm)
+
\tilde{\lambda}_t   \big)
\left(P_t  -\vartheta\right)^{-}
\,+\,
\big(\gamma_t \hatp_t
(1-\thisR)  +
 \thelambda_t\big) (P_{t}-\vartheta
  )^{+};
\lel
whereas the continuous collateralization with $\Gamma=Q=P-{\tilde{\Theta}}$ yields
 \bll{fdivbila7}
  \theb_t (P_t-\vartheta)^{+}  -
 \thebm _t (P_t-\vartheta)^{-} .\lel

\begin{rem}\rm
\label{rem:lincase-red-pred}
If $\theb=\thebm$
(case of equal collateral borrowing and lending liquidity bases),
the TVA is linear
for every collateralization scheme of the form
$P_{t} - \tilde{\Theta}_{t}  -  \Gamma_t =\varepsilon_t$,
for some exogenous\footnote{Not depending on $\tilde{\Theta}_t$,
like with continuous collateralization.}
residual exposure $\varepsilon_t$, {cf. Remark \ref{rem:lincase}.
Setting}
\bll{fdivbila2ter}
\mygb^{b}_t     =&-\left(\gamma_t \hatq_t(1-\thisRm) +\tilde{\lambda}_t \right)\varepsilon^-_t
+\ \theb_t \left( P_t -\varepsilon_t \right)
+
\left(\gamma_t \hatp_t
(1-\thisR)+\lambda_t \right)\varepsilon^+_t
,
\lel
one ends up, again similarly to \eqref{e:second}, with
\beqa\l{e:firstter}
& \tilde{\beta}^{b}_t \tilde{{\Theta}}_t= \tilde{\E}_t \Big[ \int_t^{\Ts}\tilde{{\beta}}^{b}_{s} \mygb^{b}_s   ds  \Big]
\eeqa
for the funding-adjusted discount factor
\beqa\label{e:notcredter}
\bal
&\tilde{{\beta}}^{b}_t=\exp(-\int_0^t (r_s + b_s) ds).
\eal
\eeqa
\erem
\brem \label{rem:master}
If
$\theb=\thebm=0$, the two continuous collateralization schemes of equations \eqref{fdivbila4})
and \eqref{fdivbila7}
equally collapse to
$\tilde{\Theta}=0$ and
$\tilde{\Pi}=P=\Gamma=Q$.
\erem

\subsection{Full Collateralization CSA}

Let us define the full collateralization CSA by
$Q=\Gamma^{*}$ where $\Gamma^{*}$ is given as the solution to the following $\ff$-BSDE:
\begin{equation}\label{e:Gamma}\begin{array}{c}\left\{\begin{aligned}
&\Gamma^{*}_{\Ts}=P_{\Ts}, \mbox{and for }\ttt:\\
&d\Gamma^{*}_t-\left((r_t+\theb_t) (\Gamma^{*}_t)^{+}-(r_t+\thebm_t)  (\Gamma^{*}_t)^{-}\right)dt
= dP_t-r_tP_tdt + d\tilde{\mu}^* _t
 \end{aligned}\right.\end{array}\end{equation}
for some $\ff$-martingale $\tilde{\mu}^* $.
Then $\tilde{\Theta}:=P-\Gamma^{*}$ solves the pre-default TVA BSDE \eqref{e:predcvabsddiscintrinsic},
or in other words $\tilde{\Pi}=\Gamma^{*}.$
Indeed one has by \eqref{fdivbila1}, that for $\ttt$
\bea
\mygb_t (\Gamma^{*}_t)+r_t (P_t-\Gamma^{*}_t)       &=
 \theb_t (\Gamma^{*}_t)^{+}     -
\thebm _t (\Gamma^{*}_t)^{-},
\eea
hence
\bea
(r_t+\theb_t) (\Gamma^{*}_t)^{+}-(r_t+\thebm_t)  (\Gamma^{*}_t)^{-}    -r_tP_t        &=\mygb_t (\Gamma^{*}_t).
\eea
Therefore, the second line of \eqref{e:Gamma} reads as
\bea\begin{aligned}
-(dP_t-d\Gamma^{*}_t) &=
\mygb_t(\Gamma^{*}_t)dt-  d\tilde{\mu}^* _t=
\mygb_t(P_t-(P_t-\Gamma^{*}_t))dt-  d\tilde{\mu}^* _t,
\end{aligned}\eea
which, together with the fact that
$P-\Gamma^{*}$ vanishes at $T$,
means that $P-\Gamma^{*}$
satisfies the pre-default TVA BSDE \eqref{e:predcvabsddiscintrinsic}.

If $\theb=\thebm,$ then $\mygb_t(\tilde{\Pi}_t)+r_tP_t$ reduces to
$(r_t+b_t)\tilde{\Pi}_t$ in the BSDE
\eqref{e:pi} for the fully collateralized price $\tilde{\Pi}=\Gamma^{*}.$
 This BSDE
is thus equivalent to the following explicit expression for $\tilde{\Pi}$:
\beqa\label{e:expli}
& \tilde{\beta}^{b}_t \tilde{\Pi}_t= \tilde{\E}_t \Big[ \int_t^{\Ts}\tilde{{\beta}}^{b}_{s}dD_s  \Big],
\eeqa
{where the funding-adjusted discount factor $\tilde{\beta}^{b}$ is defined by
\eqref{e:notcredter}}. In the special case $\theb=\thebm=0,$ we have
$\Gamma^* =P$ which is the situation already considered in
Remark \ref{rem:master}. This case, {which yields $\tilde{\Theta}=0$ and
$\tilde{\Pi}=P=Q=\Gamma$}, justifies the status of formula \eqref{cleanf} as the master clean valuation formula of a fully collateralized
price at an OIS collateral funding rate $r_t.$
With such a fully collateralized CSA there is no need for pricing-and-hedging a (null) TVA. The
problem boils down to the computation of a clean price $P$ and a related hedge, see e.g. \cite{Crepey11b2} for possible clean hedge specifications.

\subsection{Pure Funding}

In case $\gamma=0$, {which {is a} no counterparty risk, pure funding issue case,} the CSA value process $Q$ plays no actual role.
In particular the $dt$-coefficient of the BSDE
\eqref{e:pi} for $\tilde{\Pi}$ is given by
$$\mygb_t(\tilde{\Pi}_t)+r_tP_t=(r_t +b_t) \Gamma^+_t- (r_t +\thebm_t) \Gamma^-_t
+(r_t +\lambda_t) (\tilde{\Pi}- \Gamma_t)^+ - (r_t +\tilde{\lambda}_t) (\tilde{\Pi}-\Gamma_t)^- ,
$$
where $\tilde{\lambda}=\bar{\lambda}$ is a pure liquidity external borrowing basis.
 In case $\Gamma=0$ and ${\lambda}=\tilde{\lambda}$ this results in the following explicit expression of $\tilde{\Pi}$:
\beqa\l{e:closed}
& \tilde{\beta}^{\lambda}_t \tilde{\Pi}_t= \tilde{\E}_t  \int_t^{\Ts}\tilde{{\beta}}^{\lambda}_{s}dD_s ,
\eeqa
with the funding-adjusted discount factor $\tilde{\beta}^{\lambda}$ {defined in }
\eqref{e:notcredquattro}. In the special case $\tilde{\lambda}=\bar{\lambda}=0$
 one recovers
the classical valuation formula \eqref{cleanf} for $\tilde{\Pi}=P$.

\subsection{Asymmetrical TVA Approach}

In practice the bank can hardly hedge her jump-to-default and therefore cannot monetize and benefit from her default unless and before it actually happens). If one wants to acknowledge this,
one} can avoid to reckon any actual benefit of the bank from her own default by letting $\rho=\mathfrak{r}=1$, {cf. \eqref{d:unxi}.
Such an asymmetrical TVA approach,
even though still bilateral, allows one to {avoid many} concerns of a general symmetrical TVA approach,
 {such as} the arbitrage issue that arises for $\mathfrak{r}<1$, the hypothetical and paradoxical benefit of the bank at her own default time,
and the puzzle for the bank of having to hedge her own jump-to-default risk in order to monetize this benefit before
her default
(see \cite{Crepey11b1} and \cite{Crepey11b2})). Indeed
in this case equation \eqref{fdivbila1}
reduces to
\bll{fdivbila1asym}
\mygb_t (P_t-\vartheta)+r_t \vartheta        =&-
\gamma_t \hatq_t(1-\thisRm)
(Q_{t}
-     \Gamma_t )^{-} \\
&  +\,
 \theb_t \Gamma_t^{+}
-\thebm _t \Gamma_t^{-}+ \thelambda_t \left(P_t -\vartheta- \Gamma_t\right)^{+}
- \thelambdam_t \left(P_t -\vartheta- \Gamma_t\right)^{-}
\\&+\gamma_t
\left( P_{t}-\vartheta -  Q_t \right),
\lel
where there is no
beneficial debt valuation adjustment
anymore (for $\rho=1$ the second line of \eqref{fdivbila1} vanishes), and where the
borrowing funding basis
${\thelambdam}_t$ is interpreted as a pure liquidity cost.

Note that in case of a pre-default CSA recovery scheme $Q=\tilde{\Pi}_{-}=P-\tilde{\Theta}_{-}$, an asymmetrical (but still bilateral) TVA approach
is equivalent to a unilateral TVA approach where the bank would simply disregard her own credit risk. One has in both cases
\bl
\mygb_t (P_t-\vartheta)&+r_t \vartheta      =
\theb_t \Gamma_t^{+}  -
 \thebm _t \Gamma_t^{-}\,+\,  \thelambda_t \left(P_t -\vartheta- \Gamma_t\right)^{+}   -
\big(\thelambdam_t + \gamma_t \hatq_t(1-\thisRm)\big) \left(P_t -\vartheta- \Gamma_t\right)^{-},
\el
where $\gamma \hatq$ is the intensity of default of the counterparty.
Moreover, in an asymmetrical TVA approach with $Q=\tilde{\Pi}_{-},$
an inspection of the related equations in \cite{Crepey11b2}
shows that
a perfect TVA hedge by the bank of an
isolated default of her counterparty
(obtained as the solution to the last equation in \cite{Crepey11b2})
in fact yields
a perfect hedge of the TVA jump-to-default risk altogether (default of the counterparty and/or the bank). This holds at least
provided {that} the hedging instrument which is used for that purpose (typically
a clean CDS on the counterparty) does not jump in value at an isolated default time of the bank,
a mild condition\footnote{In the notation of the concluding Subsection 4.4 of \cite{Crepey11b2}, this is given
by ``$\mathcal{R}^{1}_t = \tilde{\mathcal{P}}^{1}_t$
on $\tm<\tp\wedge T$.''}
satisfied in most models.

\section{Clean Valuations}\label{s:clean}

In the numerical Section \ref{s:cvacomp}, we shall resort to
two univariate short rate models, presented in Subsections \ref{s:vasi} and \ref{ss:lhw},
for TVA computations on an interest rate swap. These are Markovian pre-default TVA models in the sense of Subsection \ref{ss:markov}, with
factor process $X_t=r_t$.
Our motivation for considering two different models is twofold. {Firstly, we want} to
emphasize the fact that from an implementation point of view, the BSDE schemes that we use
for TVA computations are quite model-independent (at least the backward nonlinear regression stage,
after a forward simulation of the model in a first stage). Secondly, this allows one to assess the
TVA model risk.

\brem\label{r:multiple}
The choice of interest rate derivatives for the illustrative purpose of this paper is not innocuous.
 The basic credit risk reduced-form methodology of this paper is suitable for situations of reasonable dependence
 between the reference contract and the two parties (reasonable  or unknown, e.g. the dependence between interest rates and credit is not liquidly priced on the market, see \cite{BrigoPallavicini08}). For cases of strong dependence such as counterparty risk on credit derivatives,
 the additional tools
of \cite{Crepey13}
 are necessary.

 Talking about interest rate derivatives, one should also mention the systemic counterparty risk,
 referring to various significant spreads which emerged since August 2007 between quantities that were very similar before, like OIS swap rates and LIBOR swap rates of different tenors. Through its discounting implications, this systemic component of counterparty risk has impacted all derivative markets. This means that in the current market conditions, one should actually use multiple-curve clean value models of interest rate derivatives in the TVA computations (see for instance \cite{CrepeyGrbacNguyen11} and the references therein). {In order} not to blur the main flow of argument we postpone this to a follow-up work.
\erem

\subsection{Products}\label{s:prod}

In the sequel we shall deal with the following interest rate derivatives: forward rate agreements (FRAs), IR swaps and {caps, whose definitions we provide below. The latter are used in Section \ref{s:cvacomp} for calibration purposes.} The underlying rate for all these derivatives is the LIBOR rate. {We work under the usual convention} that the LIBOR rate is set in advance and the payments are made in arrears. As pointed out in Remark \ref{r:multiple}, we do not tackle here the multiple-curve issue. Thus, we use the classical definition of the forward LIBOR rate $L_t(T, T+\delta)$, fixed at time $t\leq T$, for the future time interval $[T, T+\delta]$:
$$
L_t(T, T+\delta) = \frac{1}{\delta} \left( \frac{B_t(T)}{B_t(T+\delta)} - 1\right),
$$
where $B_{\cdot}(T)$ denotes the time-$t$ price of a zero coupon bond with maturity $T$.

\begin{defi}\em
\label{defi:FRA}
 A forward rate agreement (FRA) is a financial contract which {fixes the interest rate} $K$ which will be applied to a future time interval. Denote by  $T > 0$ the future inception date, by $T+\delta$ the maturity of the contract, where {$\delta \geq 0$}, and by $N$ the notional amount. The payoff of the FRA at maturity $T+\delta$ is equal to
$$
P^{fra}(T+\delta; T, T+\delta, K, N) = N \delta (L_T(T, T+\delta) - K).
$$
The value at time $t \in [0, T]$ of the FRA is given by
\begin{equation}
\label{eq:FRA}
P^{fra}(t; T, T+\delta, K, N) = N (B_t(T) - \bar{K} B_t(T+\delta)),
\end{equation}
where $\bar{K} =  1 + \delta K$.
\end{defi}

\begin{defi}\em
\label{defi:swap}
An interest rate (IR) swap is a financial contract between two parties to exchange one stream of future interest payments for another, based on a specified notional amount $N$. {A fixed-for-floating swap is a swap in which fixed payments are exchanged for floating payments linked to the LIBOR rate. Denote by $T_0 \geq 0$ the inception date, by $T_1 < \cdots < T_n$, where $T_1 > T_0$, a collection of the payment dates and by $K$ the fixed rate.} {Under our sign convention
recalled in Section \ref{s:cva} that the clean price values
promised dividends
$dD_t$ from the bank to her counterparty,
the time-$t$ clean price of the swap $P_t$ for the bank when it pays the floating rate (case of the so-called receiver swap for the bank) is given by }
\begin{eqnarray}
\label{eq:swap-value-before}
 P_t=P^{sw}(t; T_1, T_n) & = &  N \left(B_t(T_0) - B_t(T_n) - K \sum_{k=1}^{n} \delta_{k-1} B_t(T_k)  \right),
\end{eqnarray}
where $t \leq T_0$ and $\delta_{k-1} = T_{k} - T_{k-1}$. The swap rate $K_t$, i.e. the fixed rate $K$ making the value of the swap at time $t$ equal to zero is given by
\begin{eqnarray}
\label{eq:swap-value}
K_t & = &   \frac{B_t(T_0) - B_t(T_n)}{\sum_{k=1}^{n} \delta_{k-1} B_t(T_k)}.
\end{eqnarray}
\end{defi}
{The value of the swap from initiation onward, i.e. the time-$t$ value, for $T_0\leq t< T_n$, of the swap is given by
\begin{eqnarray}
\label{eq:swap-value-gen}
  \nonumber P^{sw}(t; T_1, T_n) & = &  N \left( \left(\frac{1}{B_{T_{k_t -1}}(T_{k_t})}-K \delta_{k_t-1} \right) B_t(T_{k_t}) - B_t(T_n) - K \sum_{k=k_t +1}^{n} \delta_{k-1} B_t(T_k)  \right),\\
\end{eqnarray}
where $T_{k_t}$ is the smallest $T_k$ (strictly) greater than $t.$ If the bank pays the fixed rate in the swap
(case of the so-called payer swap from the bank's perspective), then
the corresponding clean price $\bar{P}_t$ is given by $\bar{P}_t=-P^{sw}(t; T_1, T_n)$}.

\begin{defi}\em
An interest rate cap (respectively floor) is a financial contract in which the buyer receives payments at the end of each period in which the interest rate exceeds (respectively falls below) a mutually agreed strike. The payment that the seller has to make covers exactly the difference between the strike $K$ and the interest rate at the end of each period. Every cap (respectively floor) is a series of caplets (respectively floorlets). The payoff of a caplet with strike $K$ and exercise date $T$, which is settled in arrears, is given by
$$
P^{cpl}(T;T, K) =  \delta\, (L_T(T, T+\delta) - K )^+.
$$
\end{defi}
The time-$t$ price of a caplet with strike $K$ and maturity $T$
is given by, with $\thebar{K}=1 +\delta K$,
\begin{eqnarray}
P^{cpl}(t;T, K) & = & \delta\, B_t(T+\delta) \E^{\P^{T+\delta}} \left[\left(L_T(T, T+\delta) - K \right)^+ \,\Big| \, \Fs_t \right]\nonumber\\
& = & B_t(T+\delta) \E^{\P^{T+\delta}} \left[ \left( \frac{1}{B_T(T+\delta)} - \thebar{K} \right)^+ \,\Big| \, \Fs_t \right]\nonumber\\
& = & \thebar{K} B_t(T) \E^{\P^{T}} \left[ \left( \frac{1}{\thebar{K}} -  B_T(T+\delta) \right)^+ \,\Big| \, \Fs_t \right]\nonumber\\
&=&  \thebar{K} \E \left[ \exp^{- \int_t^T r_s \ds} \left( \frac{1}{\thebar{K}} -  B_T(T+\delta) \right)^+ \,\Big| \, \Fs_t  \right].
\label{eq:caplet-price}
\end{eqnarray}
The next-to-last equality is due to the fact that the payoff $\left(\frac{1}{B_T(T+\delta)} - \thebar{K}\right)^+$ at time $T+\delta$ is equal to the payoff {$B_T(T+\delta) \left(\frac{1}{B_T(T+\delta)} - \thebar{K} \right)^+ = \thebar{K}\left( \frac{1}{\thebar{K}} - B_T(T+\delta)\right)^+$ } at time $T$.
The last equality is obtained by changing from the forward measure $\P^T$ to the spot martingale measure $\P$, cf. \cite[Definition 9.6.2]{MusielaRutkowski05}. The above equalities say that a caplet can be seen as a put option on a zero coupon bond.

In the two models considered in the next subsections, the counterparty clean
price $P$ of an interest rate derivative satisfies, as required for
\eqref{eq:mark},
\beqa\label{e:mar}P_t=  P(t, X_t)\eeqa
for all vanilla interest rate derivatives including IR swaps, caps/floors and swaptions.

\subsection{Gaussian Vasicek short rate model}\label{s:vasi}
In the Vasicek model the evolution of the short rate $r$ is described by the following SDE
{$$
\ud r_t = a (k-r_t) \dt +  \ud W^{\sigma}_t,
$$
where {$a, k >0$} and $W^{\sigma}$ is a Brownian motion with volatility $\sigma >0$ on the filtered probability space $(\Omega,\G_T, \ff, \mathbb{P})$.} The unique solution to this SDE is given by
$$
r_t = r_0 e^{-a t} + k (1 - e^{-a t}) +  \int_0^t e^{- a(t-u)}\ud W^{\sigma}_u.
$$
{The zero coupon bond price $B_t(T)$ in this model can be written as an exponential-affine function of the current level of the short rate $r$. One has
\begin{equation}
\label{eq:ATS}
B_t(T) = e^{m_{va}(t,T) + n_{va}(t,T) r_t},
\end{equation}
where
\begin{equation}
\label{eq:Vas-m}
m_{va}(t,T) = R_{\infty} \left( \frac{1}{a} \left(1 - e^{-a(T-t)}\right) - T +t\right)  - \frac{\sigma^2}{4 a^3}  \left(1 - e^{-a(T-t)}\right)^2
\end{equation}
with
$R_{\infty}=k - \frac{\sigma^2}{2a^2},$
and
\begin{equation}
\label{eq:Vas-n}
n_{va}(t,T) := - e^{a t} \int_t^T e^{- a u} \du = \frac{1}{a} \left( e^{-a(T-t)} - 1\right).
\end{equation}
}

The clean price $P$ for FRAs and interest rate swaps can be written as
$$
P_t=P(\rc{t,} r_t), \qquad t \in [0, T%^*
],
$$
inserting the expression \eqref{eq:ATS} for the bond price $B_t(T)$ into equations \eqref{eq:FRA}, \eqref{eq:swap-value-before} and \eqref{eq:swap-value-gen} from Definitions \ref{defi:FRA} and \ref{defi:swap}.

{In particular, the time-$t$ price, for $T_0 \leq t < T_n$, of the interest rate swap is given by
\begin{eqnarray}
\label{eq:swap-Vasicek}
\nonumber %P^{sw}(t; T_1, T_n)
P_t & =  & N \Big( \Big( e^{-(m_{va}({T_{k_t-1}},T_{k_t}) + n_{va}({T_{k_t-1}},T_{k_t}) r_{T_{k_t-1}})} - K \delta_{k_t-1}  \Big) e^{m_{va}(t,T_{k_t}) + n_{va}(t,T_{k_t}) r_t}   \\
&& \qquad - e^{m_{va}(t,T_n) + n_{va}(t,T_n) r_t} - K \sum_{k=k_t+1}^{n} \delta_{k-1}  e^{m_{va}(t,T_k) + n_{va}(t,T_k) r_t} \Big),
\end{eqnarray}
which follows from \eqref{eq:swap-value-gen} and \eqref{eq:ATS}. In the above equation $m_{va}(t, T_k)$ and $n_{va}(t, T_k)$ are given by \eqref{eq:Vas-m} and  \eqref{eq:Vas-n}.
}

\subsubsection{Caplet}\label{ss:caplet-vasi}

To price a caplet at time 0
in the Vasicek model, one uses \eqref{eq:caplet-price} with
$$
B_T(T+\delta) = e^{m_{va}(T,T+\delta) + n_{va}((T,T+\delta) r_T}
$$
for $m_{va}(T, T+\delta)$ and $n_{va}(T, T+\delta)$ given by \eqref{eq:Vas-m} and \eqref{eq:Vas-n}.
Combining this with Proposition 11.3.1 and the formula on the bottom of page 354 in \cite{MusielaRutkowski05} yields (recall $\thebar{K}=1 +\delta K$):
\beqa\label{eq:caplet-price-vasi}
P^{cpl}(0;T, K) = B_0(T) \Phi (-d_- ) - \thebar{K} B_0(T+\delta)  \Phi (-d_+ ),
\eeqa
where $\Phi$ is the Gaussian distribution function and
\bel
& d_{\pm}   = \frac{\ln \left(\frac{B_0(T+\delta)}{B_0(T)}\thebar{K}\right)  }{\Xi\sqrt{T}
}\pm  \frac{1}{2}  {\Xi}\sqrt{T}
\eel
with
\beql{e:capletvas} \Xi^2 T
 := \frac{\sigma^2}{2 a^3} \left(1 - e^{-2 a  T }\right) \left(1 - e^{- a  {\delta } }\right)^2.
\eeql

\subsection{L\'evy Hull-White short rate model}\label{ss:lhw}

In this section we recall a one-dimensional L\'evy Hull-White model obtained within the HJM framework. {Contrary to the Vasicek model}, this model fits automatically the initial bond term structure {$B_0(T)$}.

As in Example 3.5 of \cite{CrepeyGrbacNguyen11},  we consider the \lev Hull--White extended Vasicek model for the short rate $r$ given by
\begin{eqnarray}\label{e:rt}
\ud r_t =\thealpha (\kappa(t) - r_t) \dt + \ud Z^{\varsigma}_t,
\end{eqnarray}
where $\thealpha >0$ and $Z^{\varsigma}$ denotes {a L\'evy process described below}. Furthermore,
\begin{eqnarray}\label{e:rho}
\kappa(t) = f_0(t) + \frac{1}{\thealpha}  {\partial_t} f_0(t) + \psi_{\varsigma}\left(\frac{1}{\thealpha} \left( e^{-\thealpha t }- 1\right)\right) - \psi_{\varsigma}' \left(\frac{1}{\thealpha} \left( e^{-\thealpha t }-1\right) \right)\frac{1}{\thealpha} e^{-\thealpha t },
\end{eqnarray}
{where $f_0(t)= - \partial_t \log B_0(t)$} and $\psi_{\varsigma}$ denotes the cumulant function of $Z^{\varsigma}$; see
\cite[Example 3.5]{CrepeyGrbacNguyen11} with the
volatility specification $\sigma_s(T)= e^{-{\thealpha}(T-s)}$, $0\leq s\leq T $, therein.

{In this paper we shall use
an inverse Gaussian (IG) process $Z^{\varsigma}=(Z^{\varsigma}_t)_{t \geq 0}$, which is a pure-jump, infinite activity, subordinator (nonnegative \lev process), providing an explicit control on the sign of the short rates (see \cite{CrepeyGrbacNguyen11}). The IG process is obtained from a standard Brownian motion $W$ by setting
$$
Z^{\varsigma}_t = \inf \{s > 0: W_s + \varsigma s > t\},
$$
where $\varsigma > 0$. Its \lev measure is given by
$$
F_{\varsigma}(\dx) = \frac{1}{\sqrt{2 \pi x^3}} e^{-\frac{\varsigma^2 x}{2}} \indik_{\{x>0\}} \, \dx.
$$
The distribution of $Z^{\varsigma}_t$ is $IG(\frac{t}{\varsigma}, t^2)$. The cumulant function $\psi_{\varsigma}$ exists for all $z \in [- \frac{\varsigma^2}{2}, \frac{\varsigma^2}{2}]$ (actually for all $z \in (-\infty, \frac{\varsigma^2}{2}]$ since $F_{\varsigma}$ is concentrated on $(0, \infty)$)
and is given by
\begin{equation}
\label{eq:cumul-IG}
\psi_{\varsigma}(z) = \varsigma \left(1- \sqrt{1-2 \frac{z}{\varsigma^2}}\right).
\end{equation}

{Similarly to the Gaussian Vasicek model, the bond price $B_t(T)$ in the \lev Hull-White short rate model can be written as an exponential-affine function of the current level of the short rate $r$:
\begin{equation}
\label{eq:ATS-Levy}
B_t(T) = e^{m_{le}(t,T) + n_{le}(t,T) r_t},
\end{equation}
where
\begin{eqnarray}
\label{eq:Levy-m}
m_{le}(t,T) & := & \log \left( \frac{B_0(T)}{B_0(t)}\right) - n(t,T) \left[f_0(t) + \psi_{\varsigma} \left(\frac{1}{\thealpha} \left( e^{-\thealpha t } -1\right)\right) \right] \\
\nonumber && - \int_0^t \left[\psi_{\varsigma} \left(\frac{1}{\thealpha} \left( e^{-{\thealpha}(T-s)}-1\right)\right) - \psi_{\varsigma} \left(\frac{1}{\thealpha} \left( e^{-{\thealpha}(t-s)}-1\right)\right)\right] \ds
\end{eqnarray}
and
\begin{equation}
\label{eq:Levy-n}
n_{le}(t,T) := - e^{\thealpha t} \int_t^T e^{-\thealpha u} \du = \frac{1}{\thealpha} \left( e^{-{\thealpha}(T-t)} - 1\right).
\end{equation}
}

{In the \lev Hull-White model the clean price $P$ for FRAs and interest rate swaps can be written as
$$
P_t=P(\rc{t,} r_t), \qquad t \in [0, T],
$$
by combining the exponential-affine representation \eqref{eq:ATS-Levy} of the bond price $B_t(T)$ and Definitions \ref{defi:FRA} and \ref{defi:swap}.
In particular, the time-$t$ price, for $T_0 \leq t < T_n$, of the swap is given by
\begin{eqnarray}
\label{eq:swap-Levy}
\nonumber  P_t & =  & N \Big( \Big( e^{-(m_{le}({T_{k_t-1}},T_{k_t}) + n_{le}({T_{k_t-1}},T_{k_t}) r_{T_{k_t-1}})} - K \delta_{k_t-1}  \Big) e^{m_{le}(t,T_{k_t}) + n_{le}(t,T_{k_t}) r_t}   \\
&& \qquad - e^{m_{le}(t,T_n) + n_{le}(t,T_n) r_t} - K \sum_{k=k_t+1}^{n} \delta_{k-1}  e^{m_{le}(t,T_k) + n_{le}(t,T_k) r_t} \Big),
\end{eqnarray}
which follows from \eqref{eq:swap-value-gen} and \eqref{eq:ATS-Levy}. In the above equations $m_{le}(t, T_k)$ and $n_{le}(t, T_k)$ are given by \eqref{eq:Levy-m} and  \eqref{eq:Levy-n}.

\subsubsection{Caplet}\label{ss:caplet-lhw}
{To calculate the price of a caplet at time 0
in the L\'evy Hull-White model,
one can replace
$\bar{B}^*$ with $B$, $\bar{\Sigma}^*$ with $\Sigma$,
 $\bar{A}^*$ with $A$, and insert ${\Sigma}^*=0$ and ${A}^*=0$
in Subsection 4.4 of \cite{CrepeyGrbacNguyen11}, thus obtaining the time-$0$ price of the caplet
\begin{eqnarray*}
P^{cpl}(0;T, K)
& = & B_0(T + \delta) \E^{\P^{T+\delta}} \left[ \left( \frac{1}{B_T(T+\delta)} - \thebar{K} \right)^+ \, \right]\\
& = & B_0(T + \delta) \E^{\P^{T+\delta}} \left[ \left( e^{Y} - \thebar{K} \right)^+ \, \right]
\end{eqnarray*}
with
\begin{eqnarray*}
Y:= \log \frac{B_0(T)}{B_0(T+\delta)} +  \int_0^{T} (A_s(T+\delta) - A_s(T))  \ds + \int_0^{T} (\Sigma_s(T+\delta) -  \Sigma_s(T))  \ud Z^{\varsigma}_s,
\end{eqnarray*}
{where $\Sigma_s(t) = \frac{1}{\alpha}\big(1 - e^{-\alpha(t-s)}\big)$ and $A_s(t) = \psi_{\varsigma}(-\Sigma_s(t))$, for $0 \leq s \leq t$.}
The time-0 price of the caplet is now given by (cf. \cite[Proposition 4.5]{CrepeyGrbacNguyen11})
\begin{equation}
\label{eq:caplet-Fourier}
P^{cpl}(0;T, K)  = \frac{B_0(T + \delta)}{2 \pi} \int_{\R} \frac{\thebar{K}^{1+\ii v - R} M_{Y}^{T+\delta}(R - \ii v )}{(\ii v - R) (1 +  \ii v - R )} \dv,
\end{equation}
for $R > 1$ such that $M_{Y}^{T+\delta} (R) < \infty$. The moment generating function $M_{Y}^{T+\delta}$ of $Y$ under the measure $\P^{T+\delta}$ is provided by
\begin{eqnarray}
\label{eq:mom-gen-f-capl}
M_{Y}^{T+\delta} (z) & = & \exp \left( - \int_0^T \psi_{\varsigma} (-\Sigma_s(T+\delta)) \ds \right) \\
\nonumber && \qquad \times \exp \left( z \left( \log \frac{B_0(T)}{B_0(T+\delta)} + \int_0^T \left(\psi_{\varsigma} (-\Sigma_s(T+\delta)) - \psi_{\varsigma} (-\Sigma_s(T))\right) \ds \right)\right) \\
\nonumber && \qquad \times \exp \left( \int_0^T \psi_{\varsigma} \left(  (z-1) \Sigma_s(T+\delta) - z \Sigma_s(T) \right) \ds \right),
\end{eqnarray}
for $z \in \C$ such that the above expectation is finite.
}
{Alternatively, the time-0 price of the caplet can be computed as the following expectation (cf. formula \eqref{eq:caplet-price})
\begin{equation}
\label{eq:caplet-Levy}
P^{cpl}(0;T, K) =  \thebar{K} \E \left[ \exp^{- \int_t^T r_s \ds} \left( \frac{1}{\thebar{K}} -  B_T(T+\delta) \right)^+ \right],
\end{equation}
where $r$ is given by \eqref{e:rt} and $B_T(T+\delta)$ by  \eqref{eq:ATS-Levy}.
}

\subsection{Numerics}\label{ss:cleannum}

In Section \ref{s:cvacomp} we shall present TVA computations on an interest rate swap with ten years maturity,
where the bank exchanges the swap rate $K$ against a floating LIBOR at the end of each year 1 to 10.
In order to fairly assess the TVA model risk issue, this will be done in {the} Vasicek model and {the} L\'evy Hull-White model calibrated to the same data, in the sense that they share a common initial zero-bond term structure
$B^*_0 (T)$ below, and {produce} the same price for the cap with payments at years 1 to 10 struck at $K$ (hence there is the same level of Black implied volatility
in both models at the
strike level $K$). Specifically, we set
$r_0=2\%$ and
the following Vasicek parameters:
\begin{eqnarray} \notag
\begin{array}{ccc}
&a =0.25 \sp k =0.05 \sp\sigma = 0.004
\end{array}
\end{eqnarray}
with related zero-coupon rates and discount factors denoted by
$
R^*_0(T)$
and
$
B^*_0(T)= \exp (-T R_0^*(T)) .$
{It follows from \eqref{eq:ATS} and after some simple calculations,}
\begin{eqnarray*}
R_0^*(T) &=& R_{\infty} -(R_{\infty}-r_0) \frac{1}{aT} \left(1 - e^{-a T}\right)+
\frac{\sigma^2}{4 a^3 T}  \left(1 - e^{-a T}\right)^2\\
f^*_0 (T) &  = &
\partial_T \big(T R^*_0 (T)\big)
 =k + e^{-a T} \left(r_0- k\right)-\frac{\sigma^2}{2a^2} \left(1 - e^{-a T}\right)^2\\
{\partial_T} f_0^*(T) &  = & -a e^{-a T}\left(r_0- k\right)- \frac{\sigma^2}{a}
 \left(1 - e^{-aT}\right)  e^{-a T}.
 \end{eqnarray*}
An application of formula \eqref{eq:swap-value} at time 0
yields for
the corresponding swap rate the value $K=3.8859\%$.
 We choose a swap
notional of $N=310.136066$\textdollar~so that the fixed leg of the swap
is worth $100$\textdollar~at inception.

In the L\'evy Hull-White model we use
 $\alpha=a=0.25$ (same speed of mean-reversion as in the Vasicek model), an initial bond term-structure $B_0(T)$ fitted to $B^*_0(T)$ {by} using $f_0 (T)=f^*_0 (T)$ above
 in \eqref{e:rho},
  and a value of $\varsigma=17.570728$
  calibrated to the price in the above Vasicek model of the cap with payments at years 1 to 10 struck at $K$. The calibration is done by least square minimization based on the explicit formulas for caps in both models reviewed in Subsections \ref{ss:caplet-vasi} and \ref{ss:caplet-lhw}. After calibration the price of the cap in both models is 20.161\textdollar~(for the above notional $N$ yielding a value of $100$\textdollar~for the fixed leg of the swap).

The top panels of Figure \ref{f:1} show 20 paths, expectations and $2.5$/$97.5$-percentiles over 10000 paths, simulated in the two models by an Euler scheme $\hat{r}$ for the short-rate $r$ on a uniform
time grid with 200 time steps over $[0, 10]${yrs}. Note one does not see the jumps on the right panel because we
used interpolation between the points so that one can identify better the twenty paths.

\begin{figure}[htbp]
\begin{center}
\hspace*{-0.5cm}\includegraphics[width=0.5\textwidth,height=0.3\textheight]{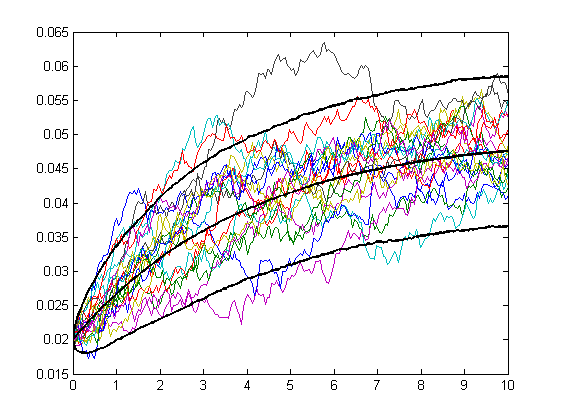}
\includegraphics[width=0.5\textwidth,height=0.3\textheight]{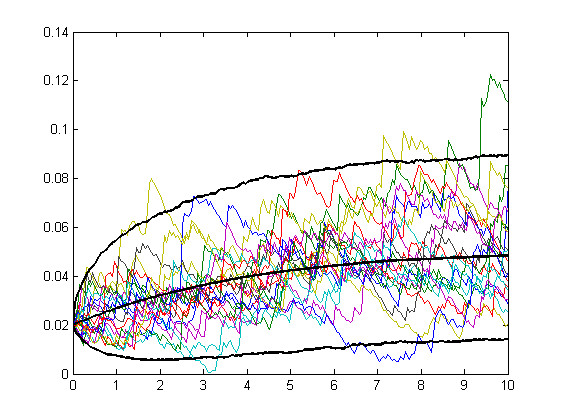}
\hspace*{-0.5cm}\includegraphics[width=0.5\textwidth,height=0.3\textheight]{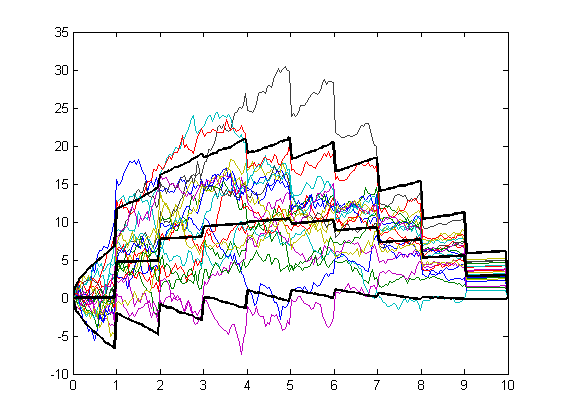}
\includegraphics[width=0.5\textwidth,height=0.3\textheight]{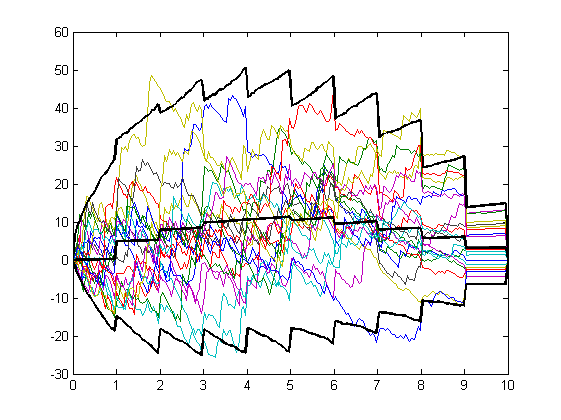}
\end{center}
\caption{20 paths with 200 time points each of the short rate $r_t$ and of the clean price process $P_t=P(t,r_t)$ of the swap. {\it Left}: Vasicek model ; {\it Right}: LHW Model.}
\label{f:1}
\end{figure}

{The top panels of Figure \ref{f:2} show the initial zero-coupon rates term structure $R^*_0 (T)$ and the corresponding forward curves
$f^*_0 (T)$, whilst the corresponding discount factors $B^*_0 (T)$ can be seen on the lower left panel.
This displays an increasing term structure of interest rates, meaning that the bank will on average be
out-of-the-money with a positive $P_t=P^{sw}_t$
in \eqref{eq:swap-value-gen}, or in-the-money with a negative $\bar{P}_t=-P^{sw}_t$,
depending on whether the bank pays floating (case of a receiver swap) or fixed rate (case of a payer swap) in the swap, see the bottom panels of Figure \ref{f:1}. Note {that} the swap price processes have quite distinct profiles in the two models {even though these are co-calibrated}.}
The bottom right panel of Figure \ref{f:2} shows the \lev Hull-White mean-reversion function $\kappa(t)$
corresponding to
$f^*_0 (T)$ through \eqref{e:rho}.

\begin{figure}[htbp]
\begin{center}
\includegraphics[width=7cm,height=5cm]{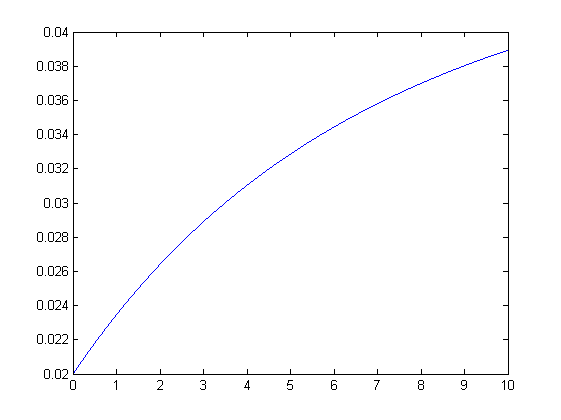}
\includegraphics[width=7cm,height=5cm]{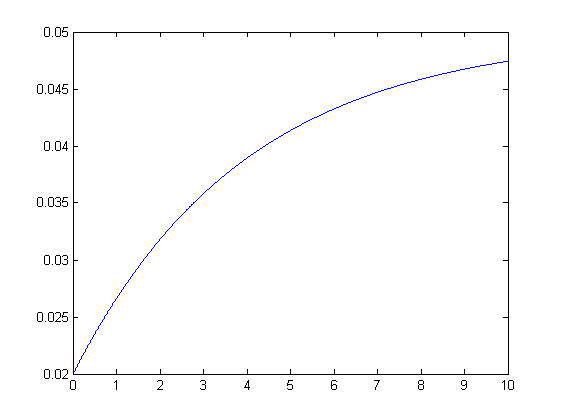}
\includegraphics[width=7cm,height=5cm]{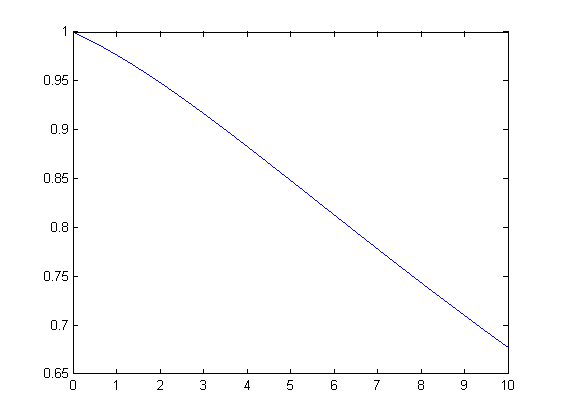}
\includegraphics[width=7cm,height=5cm]{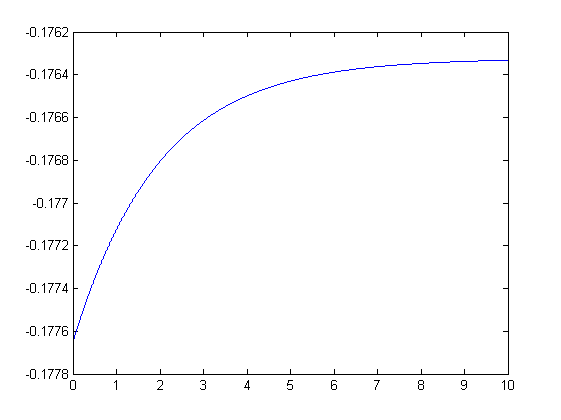}
\end{center}
\caption{{\it Top Left}: $R^*_0 (T)$ ; {\it Top Right}: $f^*_0 (T)$;
{\it Bottom Left}: $B^*_0 (T)$ ; {\it Bottom Right}: $\kappa(t)$.}
\label{f:2}
\end{figure}

\section{TVA Computations}
\label{s:cvacomp}

In this section we show with practical examples
how the CVA, DVA, LVA and RC terms {defined in} Section \ref{s:cva}
can be computed for various CSA specifications of
the coefficient $\hat{g}$ in Section \ref{s:csa}. {The computation are done for an IR swap
in the two models of Section \ref{s:clean}.}

\subsection{TVA Equations}
The generator $\mathcal{X}$ of the Gaussian Vasicek short rate $r$ in {the pre-default pricing PDE \eqref{theVI} is given by}
\bea
\bal
\mathcal{X}\tilde{\Theta}  \,(t,r)=&(a (k - r ))\partial_r \tilde{\Theta}  + \frac{1}{2} \sigma^2   \partial^2_{r^2}  \tilde{\Theta}.
\eal
\eea
Assuming \r{eq:mark}, the corresponding {TVA} Markovian BSDE writes:
$\tilde{\Theta}(\Ts,r_{\Ts})=0,$ and for $\ttt:$
\begin{eqnarray} \l{e:tvavasi}
\bal
-d\tilde{\Theta}(t,r _t)=&  \mygh(t ,r_t,\tilde{\Theta}(t,r_t)) dt  - \partial_r \tilde{\Theta}(t,r_t))  \ud W^{\sigma}_t.
\eal\end{eqnarray}

Similarly, the generator $\mathcal{X}$ of the \lev Hull-White short rate $r$ driven by an IG process is given by
{\beqa\label{eq:genlev}
\bal
\mathcal{X}\tilde{\Theta}  \,(t,r)=&(\alpha (\kappa(t) - r ))\partial_r \tilde{\Theta}  +
  \int_{\deltar >0}\Big(\tilde{\Theta}(t,  r + \deltar )
- \tilde{\Theta}(t,r )\Big)F_{\varsigma}(d\deltar),
\eal
\eeqa}where $F_{\varsigma}$ stands for the \lev measure of $Z^{\varsigma}$.\footnote{The integral converges in \r{eq:genlev} under technical conditions stated in \cite{CrepeyGrbacNguyen11}.} Assuming \r{eq:mark}, the corresponding {TVA} Markovian BSDE writes:
$\tilde{\Theta}(\Ts,r_{\Ts})=0,$ and for $\ttt:$
{\begin{eqnarray*}
\bal
-d\tilde{\Theta}(t,r _t)=&  \mygh(t ,r_t,\tilde{\Theta}(t,r_t)) dt  -
 \int_{\deltar  >0} \Big(\tilde{\Theta}(t,  r _{t-}+ \deltar
)
- \tilde{\Theta}(t,r_{t-})\Big)N^{\varsigma}(dt,d\deltar ),
\eal\end{eqnarray*}
}
where
$N^{\varsigma}$
stands for the compensated jump measure of $Z^{\varsigma}$.

\subsection{BSDE Scheme}

Even though {finding deterministic solutions of the corresponding PDEs} would be possible
in the above univariate setups, {in this paper we nevertheless favor BSDE schemes}, as they are more generic -- in real-life higher-dimensional applications deterministic schemes cannot be used anymore.
We solve \eqref{e:tvavasi} {and} \eqref{eq:genlev} by
backward regression over the time-space grids generated in Subsection
 \ref{ss:cleannum}; see the top panels of Figure \ref{f:1}.
We thus
 approximate $\tilde{\Theta}_t(\omega)$ in \eqref{e:tvavasi} and \eqref{eq:genlev} by $
 \hat{\Theta}_i^j$ on the corresponding time-space grid, where
 the time-index $i$ runs from $1$ to $n=200$ and the space-index $j$
 runs from $1$ to $m=10^4$. Denoting by $\hat{\Theta}_i= (\hat{\Theta}_i^j)_{1\le j\le m}$ the vector of TVA values on the space grid at time $i,$ we have $\hat{\Theta}_n=0$,
 and then for every $i=n-1, \cdots , 0$ and $j=1,\cdots, m$
 $$
 \hat{\Theta}_i^j= \hat{\mathbb{E}}_{i}^{j} \left(  \hat{\Theta}_{i+1} +\mygh_{i+1}\big(t ,\hat{r}_{i+1} , \hat{\Theta}_{i+1})\big)h
  \right)
 $$
for the time-step $h=\frac{T}{n}=0.02$y (one week, roughly). The conditional expectations in space at every time-step are computed by a $q$-nearest neighbor average non-parametric regression estimate
(see, e.g., \cite{HastieTibshiraniFriedman09}), with $q=5$ in our numerical experiments below.

\subsection{Numerics}

We set the following TVA parameters:
{$\gamma=10\% ,\,\theb=\thebm =\thelambda=1.5\%
   ,\, \thelambdam=4.5\%
  ,\,\hatp=  50 \% ,\,\hatq= 70\%$}
and we consider five possible CSA specifications {in this order}:
\begin{eqnarray} \notag
\begin{array}{llll}
&(\mathfrak{r},\rho,\bar{\rho})=  (40, 40,40)\%  ,\,
&Q=P  ,\,
& \Gamma=0\\
&(\mathfrak{r},\rho,\bar{\rho})=  (100,40,40)\%  ,\,
&Q=P  ,\,
& \Gamma=0  \\
&(\mathfrak{r},\rho,\bar{\rho})= (100,100,40)\%  ,\,
&Q=P  ,\,
& \Gamma=0\\
&(\mathfrak{r},\rho,\bar{\rho})= (100,100,40)\%  ,\,
&  Q=\Pi  ,\,
& \Gamma=0
\\
&(\mathfrak{r},\rho,\bar{\rho})=  (100,40,40)\%  ,\,
&Q=P  ,\,
&   \Gamma=Q=P.
\end{array}
\end{eqnarray}
Note that under the first CSA specification,
one has $\tilde{\thelambda}=4.5\%-0.6 \times 0.5 \times 10\%=1.5\%=\thelambda,$
so this is a linear TVA special case
of Remark \ref{rem:lincase-red-clean},
where the TVA at time 0 can be {computed
through a straight Monte Carlo simulation}.

Moreover, we shall study the {TVA} in the two co-calibrated Vasicek and \lev models,
and for the receiver and payer swaps. We thus consider twenty
cases (5 CSA specifications $\times$ 2 models $\times$ receiver versus payer swap).

Table \ref{t:1} shows the time-0 TVAs and the corresponding CVA/DVA/LVA/RC decompositions (four terms on the right-hand side of \eqref{nicev}) in each of the twenty cases. For benchmarking the numerical BSDE results we display in
Figure \ref{f:CI} the time-0 TVA BSDE value versus the TVA Monte Carlo mid- and $95\%$-lower and upper bounds in the {first
 CSA specification (linear)}. In all four cases the BSDE time-0 value of the TVA is close to the middle of the
confidence interval.
\begin{figure}[htbp]
\begin{center}
\hspace{1.5cm}\includegraphics[width=0.21\textwidth,height=0.13\textheight]{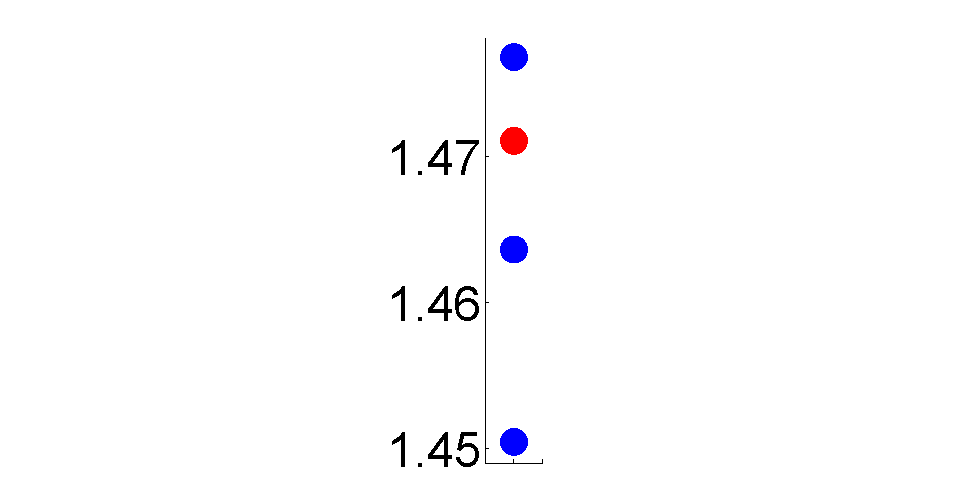}
\includegraphics[width=0.21\textwidth,height=0.13\textheight]{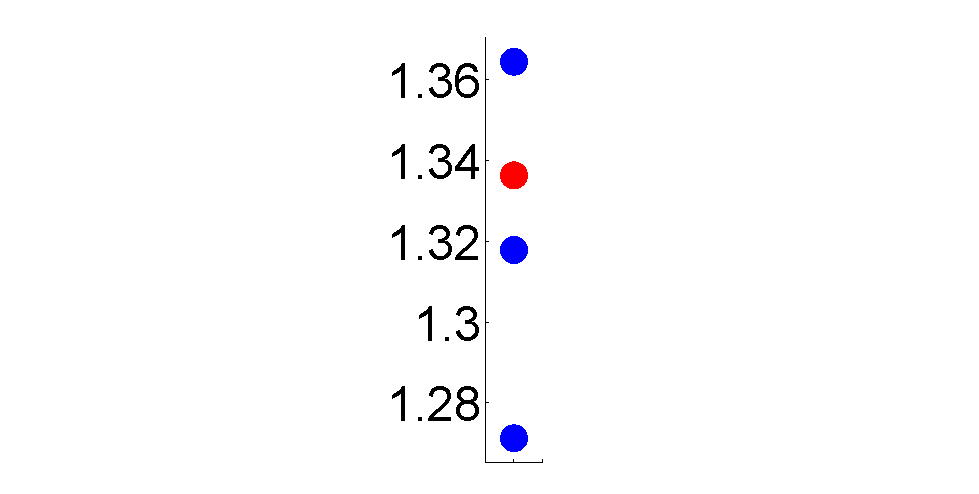}
\includegraphics[width=0.21\textwidth,height=0.13\textheight]{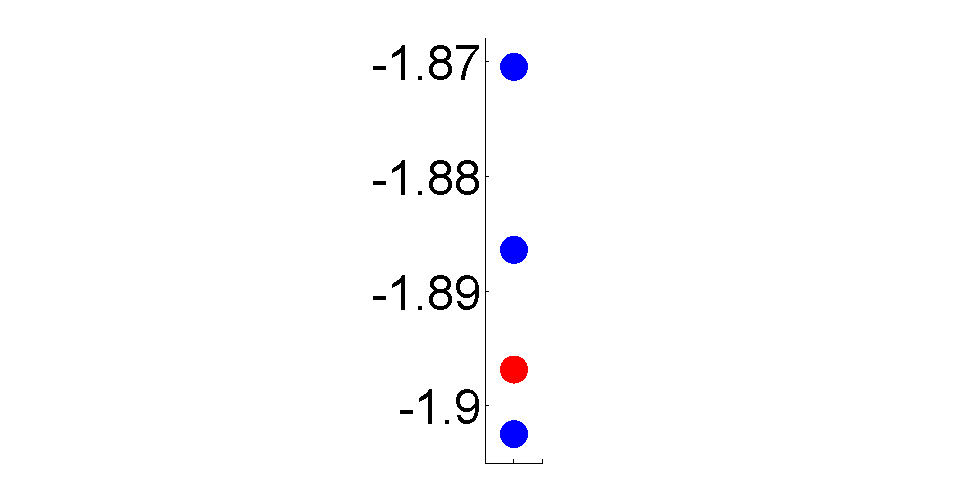}
\includegraphics[width=0.21\textwidth,height=0.13\textheight]{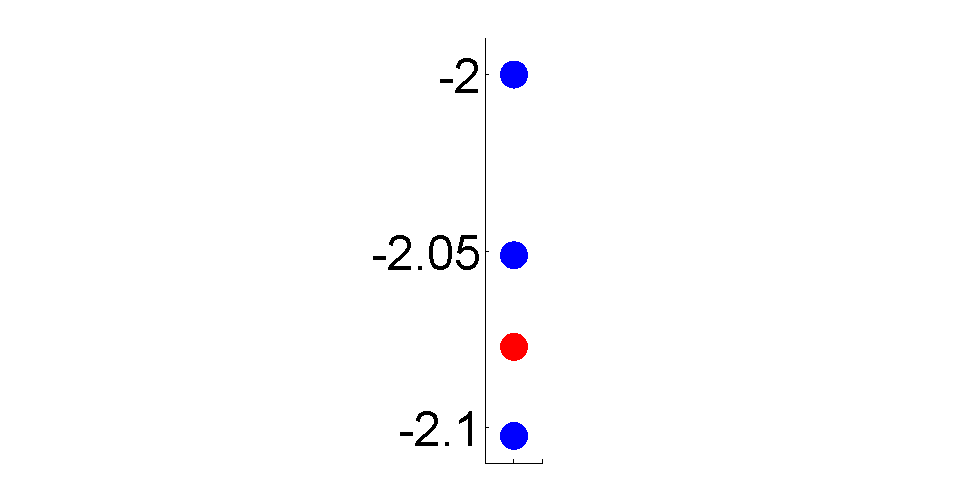}
\end{center}
\caption{Time-0 linear TVA BSDE value versus Monte Carlo mid-value and $95\%$-bounds.
\emph{From left to right}: Receiver swap in the Vasicek model; Receiver swap in the \lev model;
Payer swap in the Vasicek model; Payer swap in the \lev model.}
\label{f:CI}
\end{figure}

The numbers of Table \ref{t:1} are fully consistent with the CVA/DVA/LVA/RC interpretation of the four terms {in} the TVA decomposition {on the right-hand side} of \eqref{nicev}, given the increasing term structure of the data discussed in Subsection \ref{ss:cleannum}. For instance,
a ``high'' DVA of 1.75 in the Vasicek model and 2.34 in the \lev model
for the receiver swap in the first row of Table \ref{t:1}, is consistent with the fact that
with an increasing term structure of rates,
the bank is on average
out-of-the-money on the receiver swap with a positive $P_t=P^{sw}_t$ (see
%the bottom panels of
Figure \ref{f:1}). The CVA, on the contrary, is moderate,
as it should be for a receiver swap in an increasing term structure of interest rates, and higher (``more negative'') in the \lev than in the Vasicek model
(-0.90 versus -0.06).
The numbers of Table \ref{t:1} are not negligible at all in view of the initial value of
 $100$\textdollar~of the fixed leg of the swap. In particular, the LVA terms are quite significant
in case of the payer swap with {$\mathfrak{r}$} and/or $\rho=100\%$, see the corresponding terms {in} rows 3 and 4 {in the two bottom parts} of Table \ref{t:1}. The choice of $\mathfrak{r}$ and $\rho$
thus has tangible operational consequences, in regard of
 the ``deal facilitating''(``deal hindering'') interpretation of the
  positive (negative) TVA terms as explained {at} the end of Subsection \ref{ss:wings}. {It is worthwhile noting that all this happens} in a simplistic toy model of TVA, {in which credit risk is independent from interest rates}. These numbers could be even much {higher (in absolute value)} in a model accounting for potential wrong way risk dependence effects between {interest rates and credit risk}, {see Remark \ref{r:multiple}.}

       \begin{table}[htbp]
     \begin{center}
\begin{tabular}{|r||r|r|r|r|}
\hline
TVA & CVA & DVA & LVA & RC\\
\hline
1.47 & -0.06 & 1.75 & 0.71 & -0.92\\
\hline
1.40 & -0.06 & 1.75 & 0.64 & -0.91\\
\hline
0.40 & -0.06 & 0.00 & 0.76 & -0.29\\
\hline
0.66 & -0.08 & 0.00 & 0.74 & 0.00\\
\hline
0.43 & 0.00 & 0.00 & 0.72 & -0.29\\
\hline
\hline
         TVA & CVA & DVA & LVA& RC \\
        \hline
-1.90 & -2.45 & 0.04 & -0.68 & 1.17\\
\hline
-2.64 & -2.45 & 0.04 & -1.92 & 1.67\\
\hline
-2.67 & -2.45 & 0.00 & -1.92 & 1.68\\
\hline
-3.59 & -1.77 & 0.00 & -1.83 & 0.00\\
\hline
-0.50 & 0.00 & 0.00 & -0.81 & 0.31\\
\hline
\end{tabular}
\begin{tabular}{|r||r|r|r|r|}
\hline
TVA & CVA & DVA & LVA & RC\\
\hline
1.34 & -0.90 & 2.34 & 0.72 & -0.85\\
\hline
0.93 & -0.90 & 2.34 & 0.15 & -0.68\\
\hline
-0.45 & -0.90 & 0.00 & 0.32 & 0.12\\
\hline
-0.43 & -0.76 & 0.00 & 0.32 & 0.00\\
\hline
0.44 & 0.00 & 0.00 & 0.72 & -0.29\\
\hline
\hline
TVA & CVA & DVA & LVA & RC\\
\hline
-2.08 & -3.28 & 0.64 & -0.66 & 1.25\\
\hline
-3.17 & -3.28 & 0.64 & -2.41 & 1.92\\
\hline
-3.59 & -3.28 & 0.00 & -2.38 & 2.11\\
\hline
-4.80 & -2.49 & 0.00 & -2.26 & 0.00\\
\hline
-0.51 & 0.00 & 0.00 & -0.81 & 0.31\\
\hline
\end{tabular}
      \caption{Time-0 TVA and its decomposition.
      {\it Top}: Receiver swap ; {\it Bottom}: Payer swap. {\it Left}: Vasicek model; \textit{Right}: LHW Model.}
      \label{t:1}
      \end{center}
      \end{table}

{Figures \ref{f:4a} and \ref{f:4b} (receiver swap in the Vasicek and \lev model, respectively) and Figures \ref{f:4c} and \ref{f:4d} (payer swap in the Vasicek and \lev model, respectively)} show the ``expected exposures'' of the four right-hand side terms of the ``local'' TVA decompositions \eqref{fdivbila1}
with $\vartheta$ replaced by $\tilde{\Theta}_t$ therein. {These exposures are computed as space-averages over $10^4$ paths as a function of time $t$.}  Each time-0 integrated term of the TVA
in Table \ref{t:1} corresponds to the surface under the corresponding curve in Figures \ref{f:4a} to \ref{f:4d} (with mappings between, respectively: Figure \ref{f:4a} and the upper left corner of Table \ref{t:1}, Figure \ref{f:4b} and the upper right corner of Table \ref{t:1}, Figure \ref{f:4c} and the lower left corner of Table \ref{t:1}, Figure \ref{f:4d} and the lower right corner of Table \ref{t:1}).

\begin{figure}[htbp]
%\vspace{-4cm}
\begin{center}
\hspace*{-1cm}\includegraphics[width=1.1\textwidth,height=0.8\textheight]{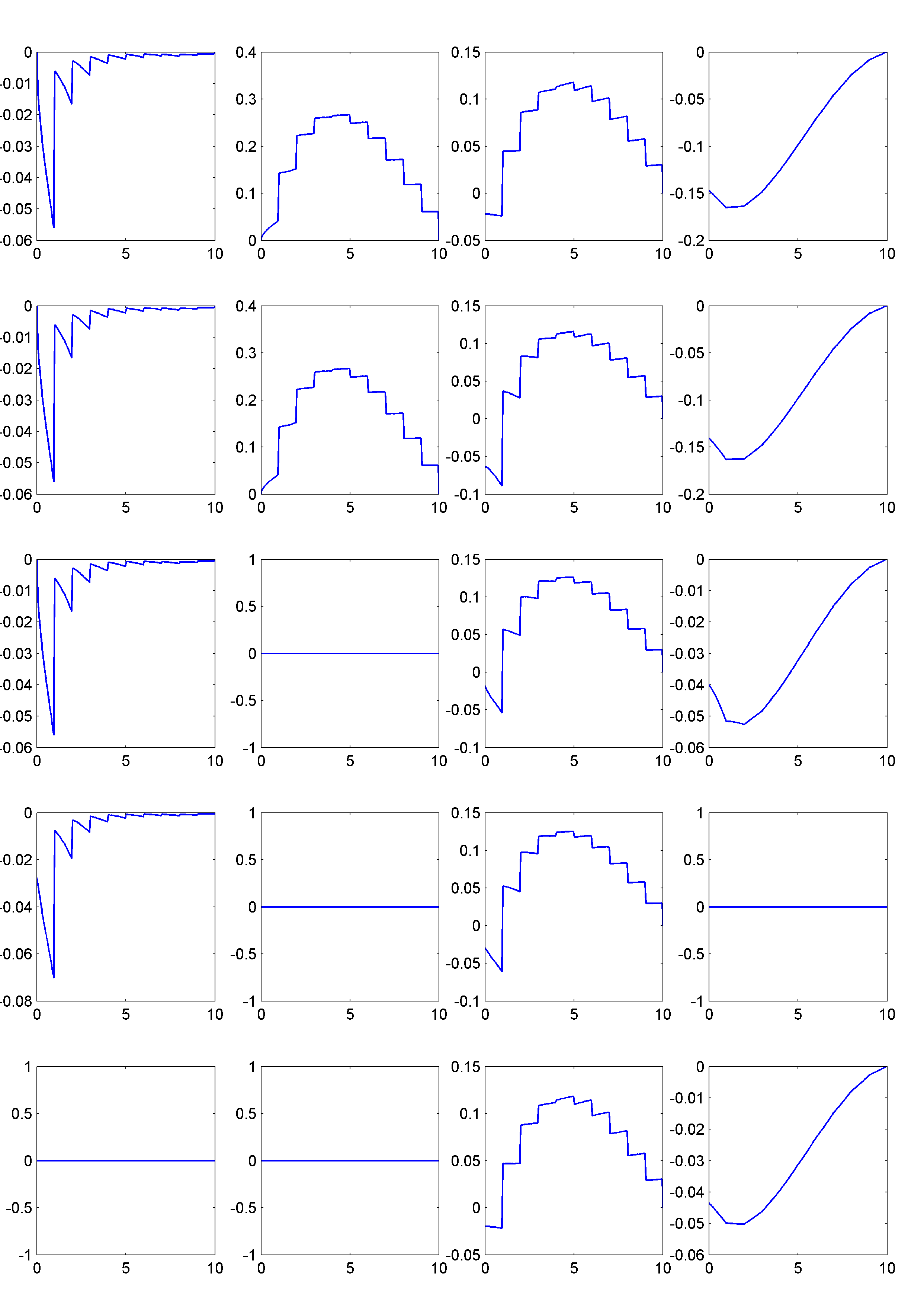}
\end{center}
\caption{Receiver swap in the Vasicek model. {\it Columns}: CVA/DVA/LVA/RC, the 4 wings of the TVA; \emph{Rows}: 5 CSA Specifications.}
\label{f:4a}
\end{figure}
\begin{figure}[htbp]
%\vspace{-4cm}
\begin{center}
\hspace*{-1cm}\includegraphics[width=1.1\textwidth,height=0.8\textheight]{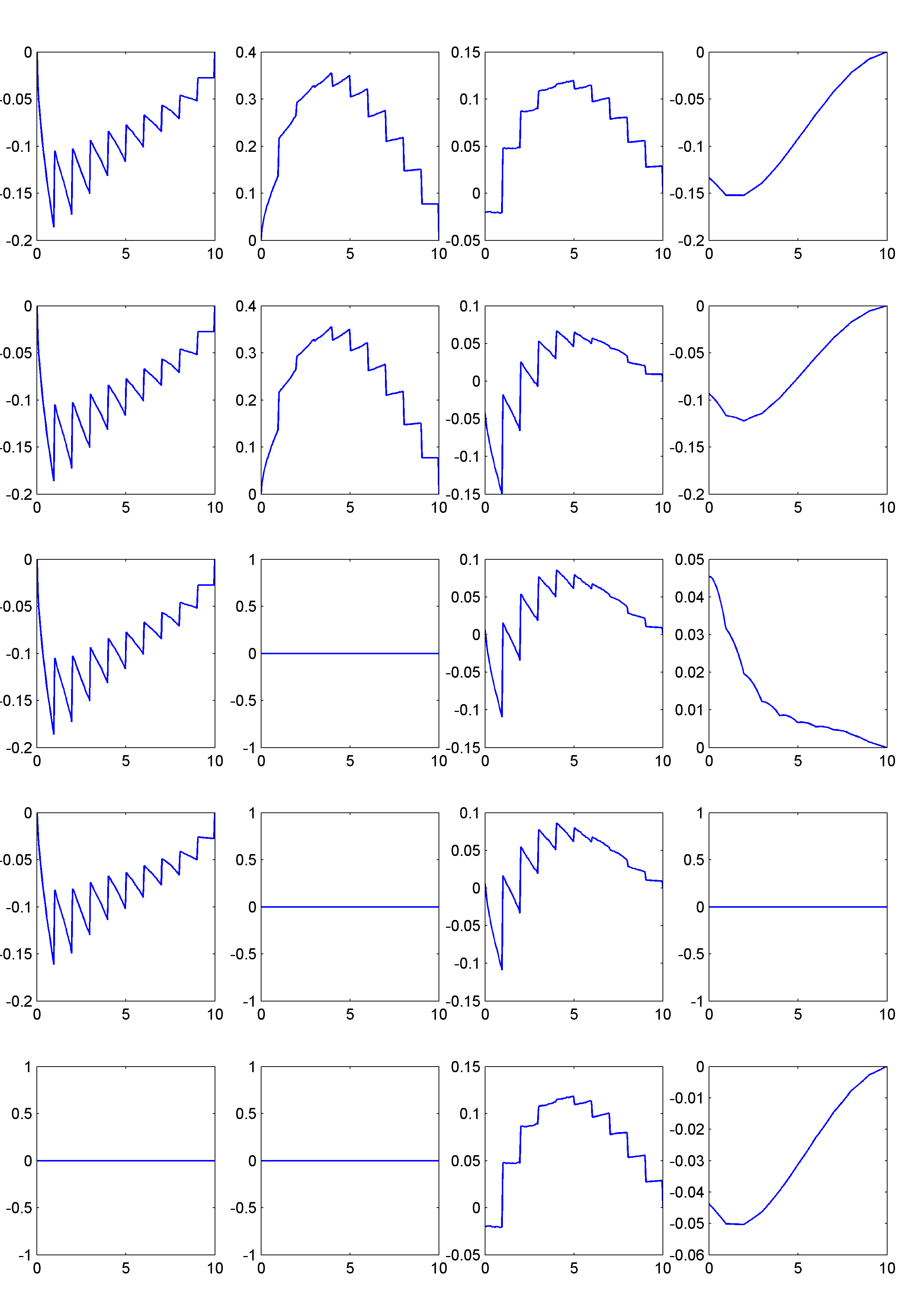}
\end{center}
%\vspace{-1cm}
\caption{Receiver swap in the \lev model. {\it Columns}: CVA/DVA/LVA/RC, the 4 wings of the TVA; \emph{Rows}: 5 CSA Specifications.}
\label{f:4b}
\end{figure}
\begin{figure}[htbp]
%\vspace{-4cm}
\begin{center}
\hspace*{-1cm}\includegraphics[width=1.1\textwidth,height=0.8\textheight]{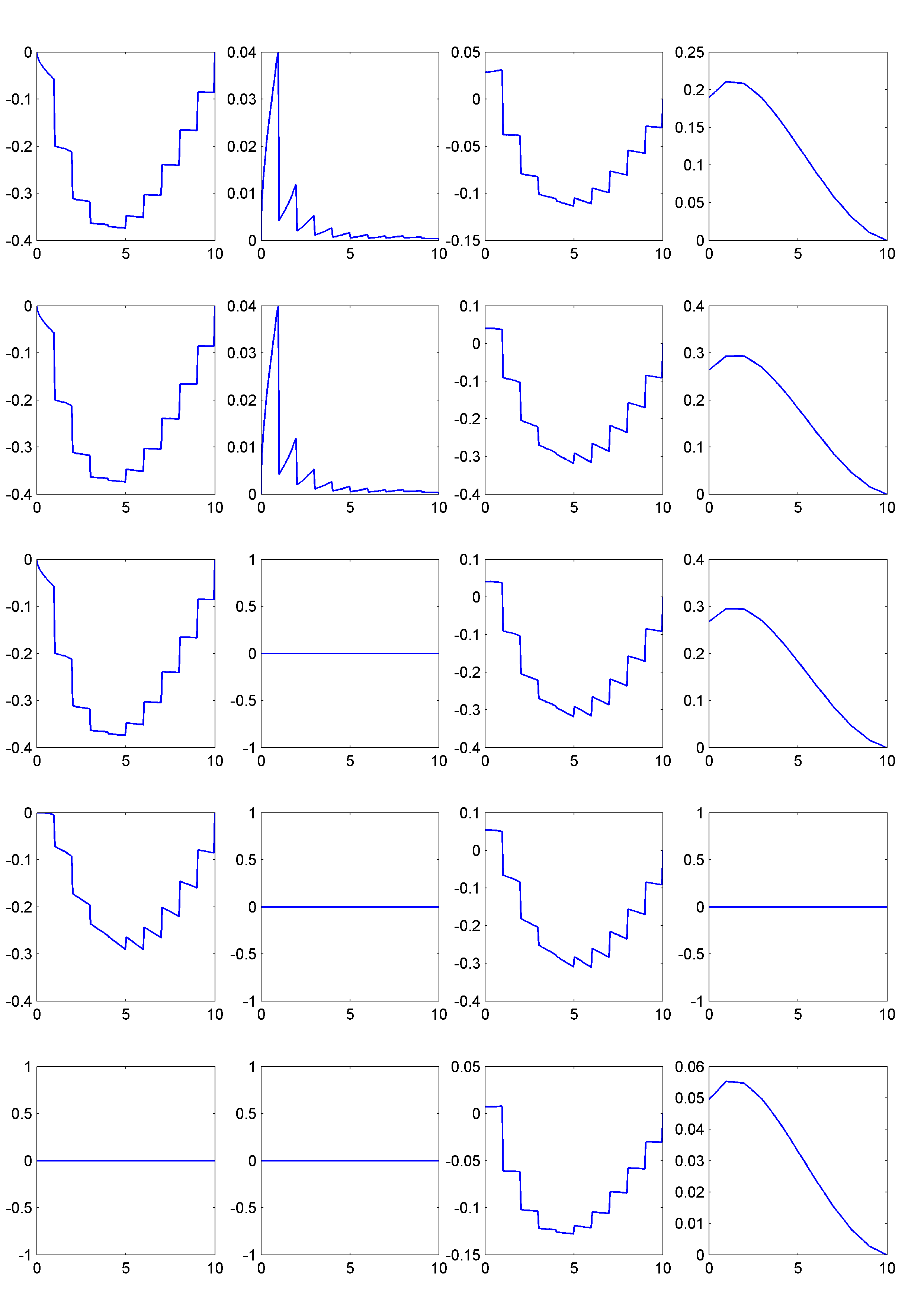}
\end{center}
%\vspace{-1cm}
\caption{Payer swap in the Vasicek model. {\it Columns}: CVA/DVA/LVA/RC, the 4 wings of the TVA; \textit{Rows}: 5 CSA Specifications.}
\label{f:4c}
\end{figure}
\begin{figure}[htbp]
%\vspace{-4cm}
\begin{center}
\hspace*{-1cm}\includegraphics[width=1.1\textwidth,height=0.8\textheight]{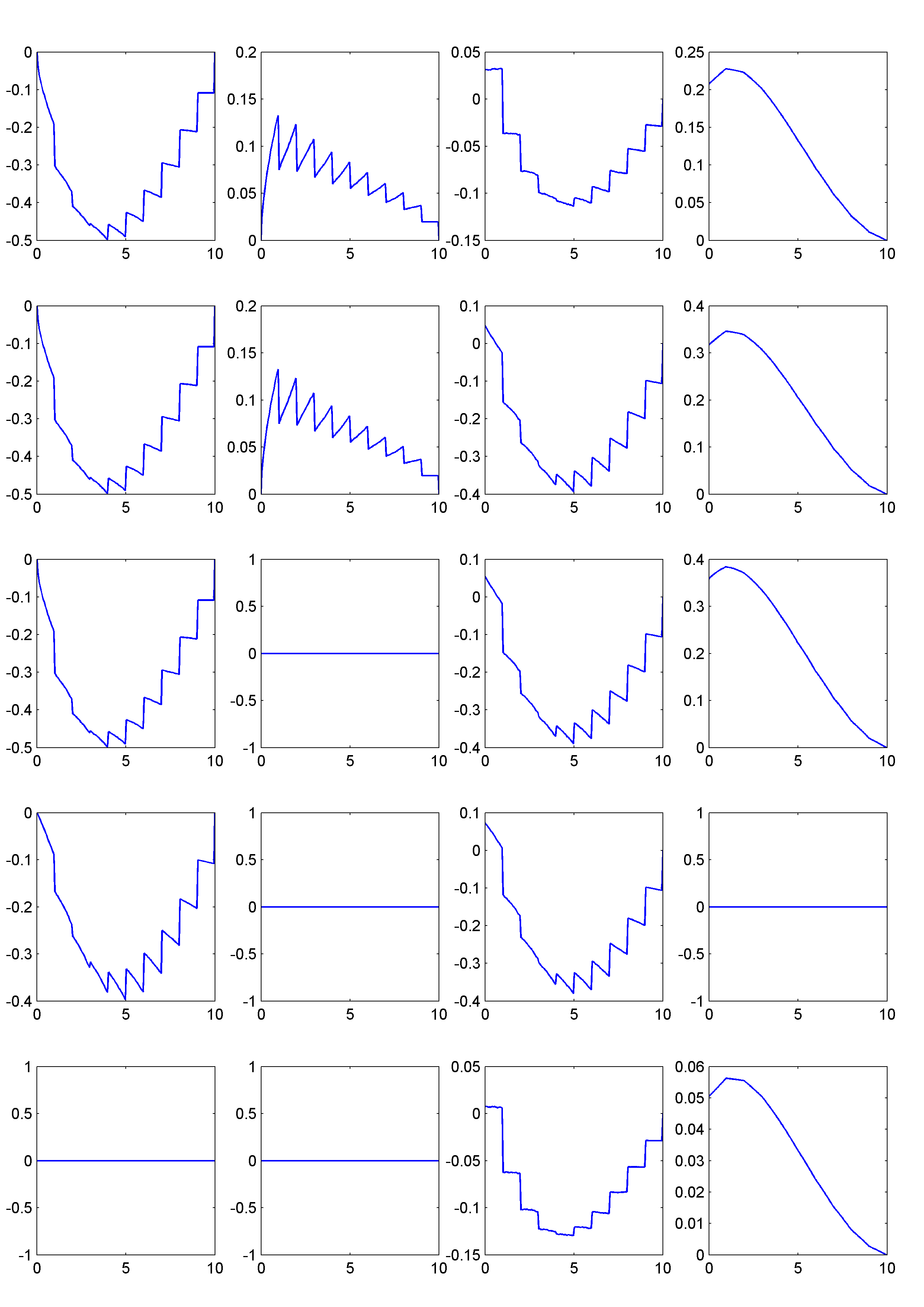}
\end{center}
%\vspace{-1cm}
\caption{Payer swap in the \lev model. {\it Columns}: CVA/DVA/LVA/RC, the 4 wings of the TVA; {\it Rows}: 5 CSA Specifications.}
\label{f:4d}
\end{figure}

Finally,
Figures \ref{f:3a} (receiver swap) and \ref{f:3b} (payer swap)
show the TVA processes in the same format as the swap clean prices {at} the bottom
of Figure \ref{f:1}.

\begin{figure}[htbp]
\vspace{-1cm}
\begin{center}
\hspace*{-0.5cm}\includegraphics[width=0.5\textwidth,height=0.18\textheight]{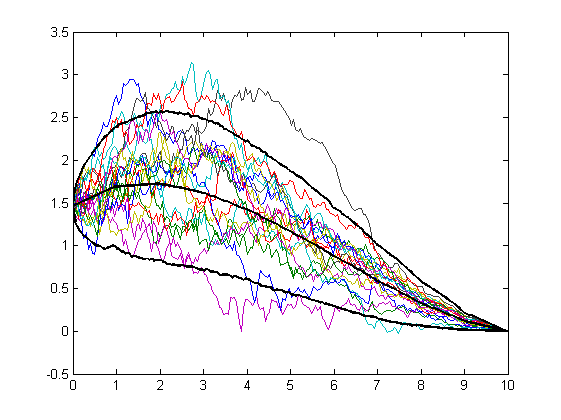}
\includegraphics[width=0.5\textwidth,height=0.18\textheight]{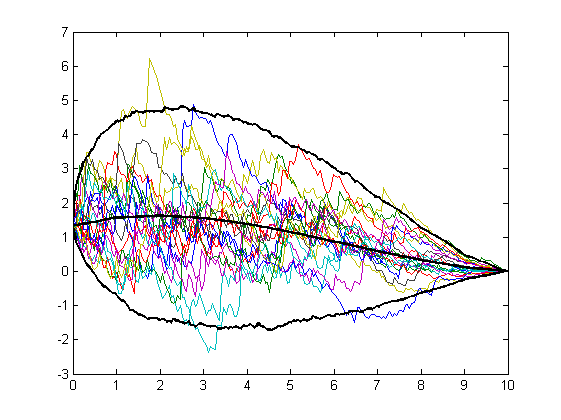}
\hspace*{-0.5cm}\includegraphics[width=0.5\textwidth,height=0.18\textheight]{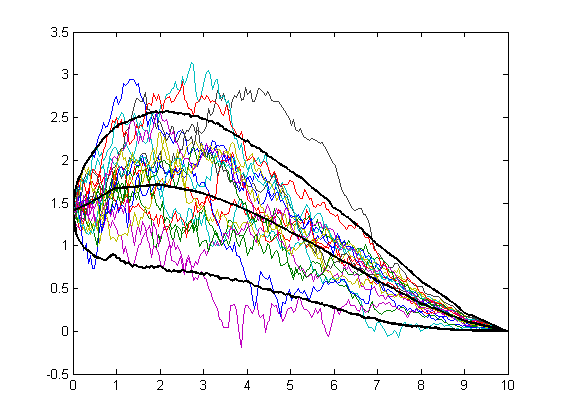}
\includegraphics[width=0.5\textwidth,height=0.18\textheight]{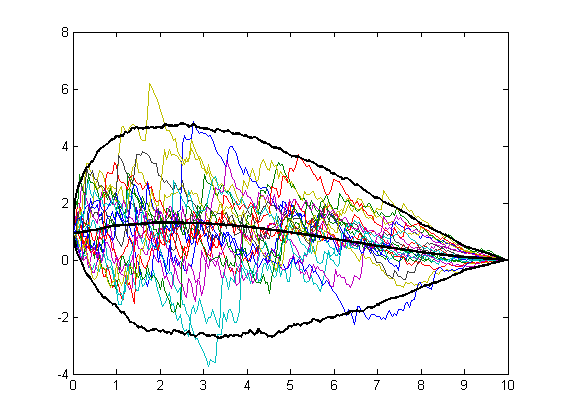}
\hspace*{-0.5cm}\includegraphics[width=0.5\textwidth,height=0.18\textheight]{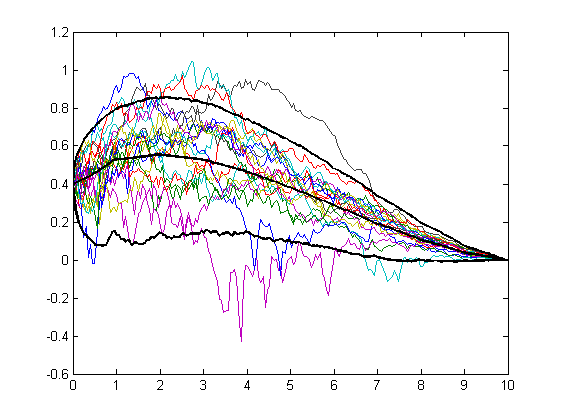}
\includegraphics[width=0.5\textwidth,height=0.18\textheight]{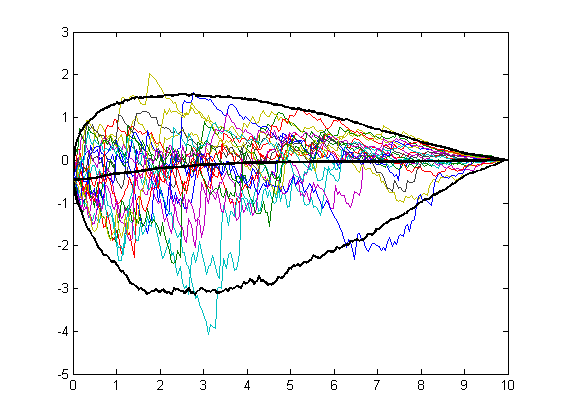}
\hspace*{-0.5cm}\includegraphics[width=0.5\textwidth,height=0.18\textheight]{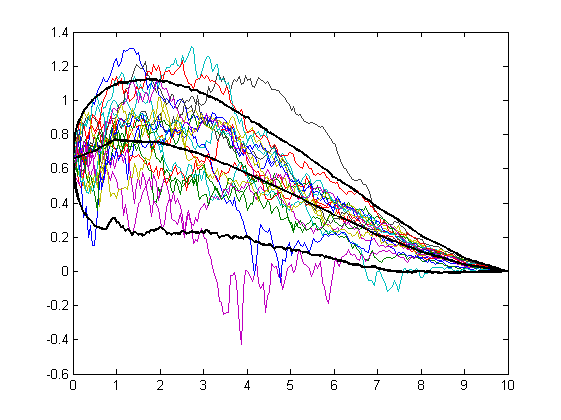}
\includegraphics[width=0.5\textwidth,height=0.18\textheight]{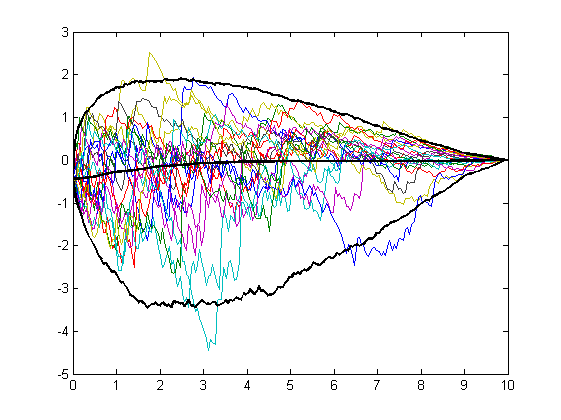}
\hspace*{-0.5cm}\includegraphics[width=0.5\textwidth,height=0.18\textheight]{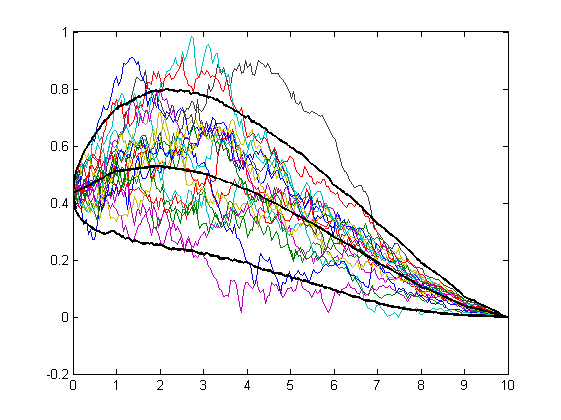}
\includegraphics[width=0.5\textwidth,height=0.18\textheight]{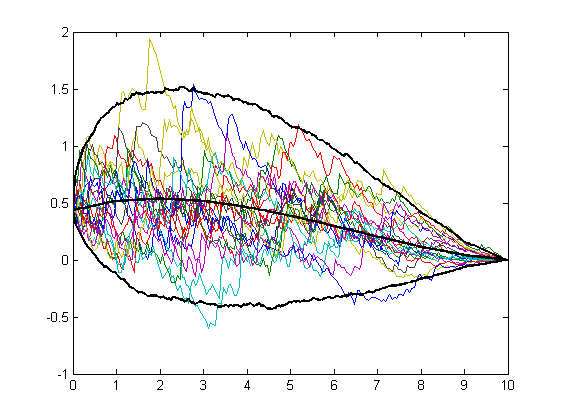}
\end{center}
\caption{Receiver swap: 20 paths with 200 time points each of the TVA process $\tilde{\Theta}_t=\tilde{\Theta}(t,r_t)$. {\it Left}: Vasicek model ; {\it Right}: LHW Model.
{\it Top to Bottom}: 5 CSA Specifications.}
\label{f:3a}
\end{figure}

\begin{figure}[htbp]
\vspace{-1cm}
\begin{center}
\hspace*{-0.5cm}\includegraphics[width=0.5\textwidth,height=0.18\textheight]{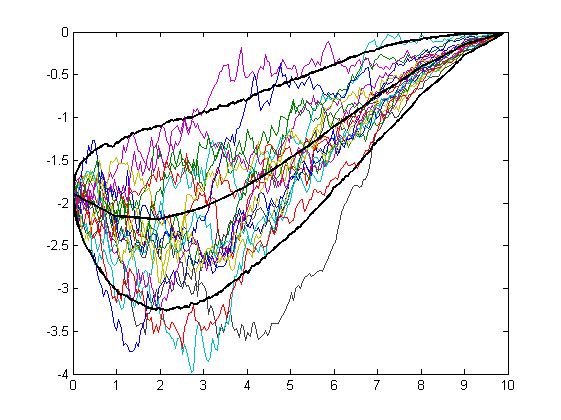}
\includegraphics[width=0.5\textwidth,height=0.18\textheight]{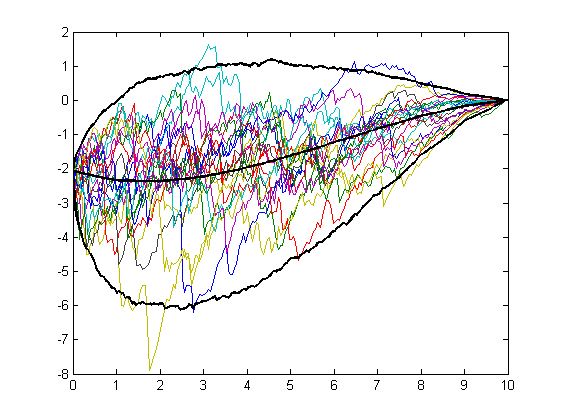}
\hspace*{-0.5cm}\includegraphics[width=0.5\textwidth,height=0.18\textheight]{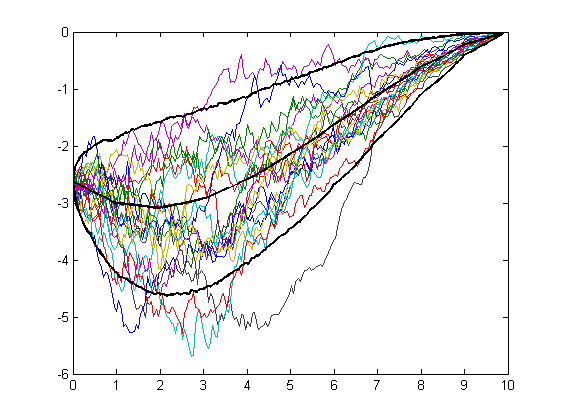}
\includegraphics[width=0.5\textwidth,height=0.18\textheight]{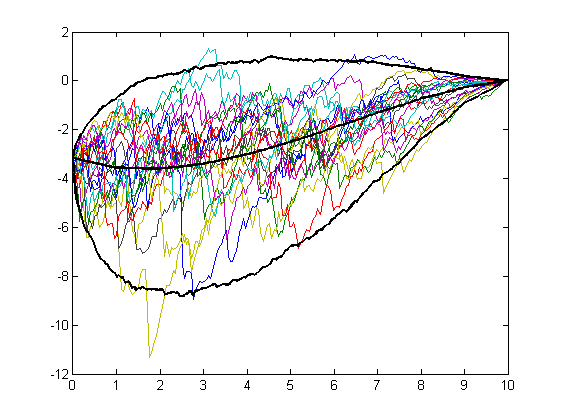}
\hspace*{-0.5cm}\includegraphics[width=0.5\textwidth,height=0.18\textheight]{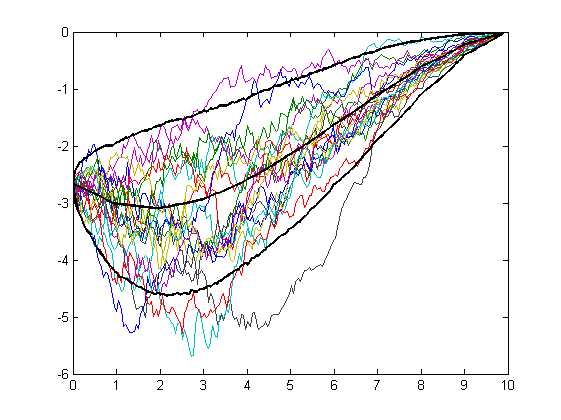}
\includegraphics[width=0.5\textwidth,height=0.18\textheight]{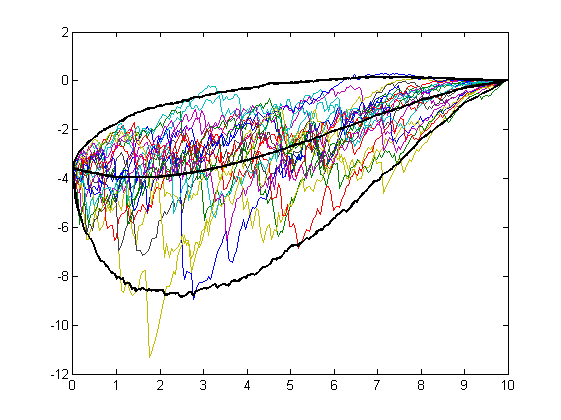}
\hspace*{-0.5cm}\includegraphics[width=0.5\textwidth,height=0.18\textheight]{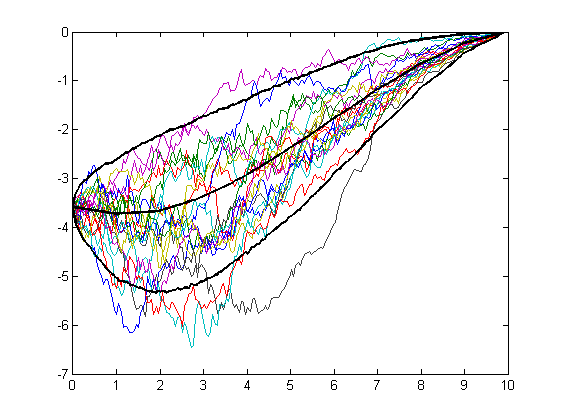}
\includegraphics[width=0.5\textwidth,height=0.18\textheight]{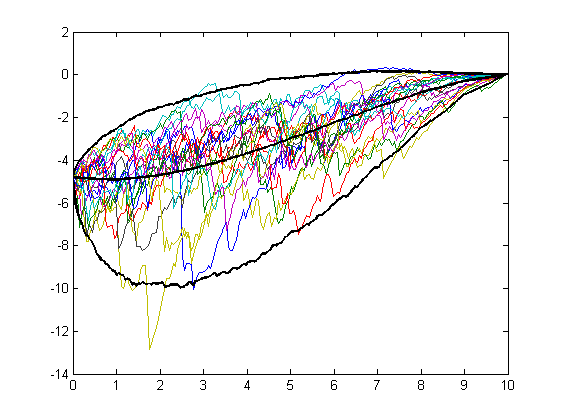}
\hspace*{-0.5cm}\includegraphics[width=0.5\textwidth,height=0.18\textheight]{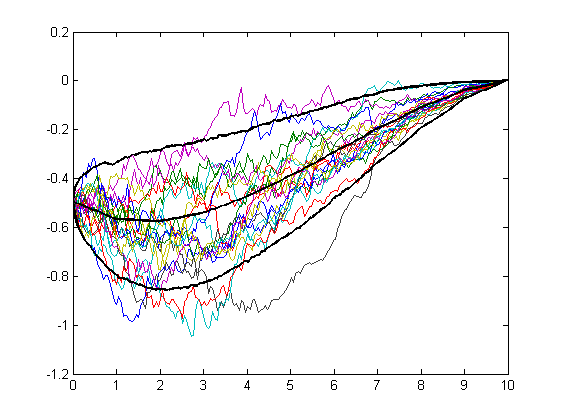}
\includegraphics[width=0.5\textwidth,height=0.18\textheight]{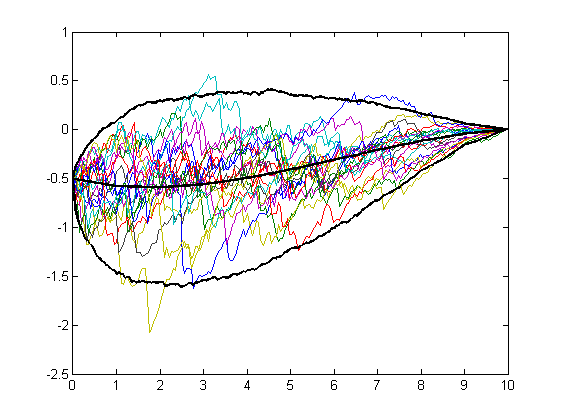}
\end{center}
\caption{Payer swap: 20 paths with 200 time points each of the TVA process $\tilde{\Theta}_t=\tilde{\Theta}(t,r_t)$. {\it Left}: Vasicek model ; {\it Right}: LHW Model.
{\it Top to Bottom}: 5 CSA Specifications.}
\label{f:3b}
\end{figure}

\section*{Conclusion}

In this paper, which is a numerical companion to
\cite{Crepey11b1} and \cite{Crepey11b2},
we show on the standing example of an interest-rate swap
how CVA, DVA, LVA and RC, the four ``wings'' (or pillars) of the TVA,
can be computed for various CSA and model specifications.
 Positive terms {such as} the DVA (resp. negative terms {such as} the CVA) can be considered as ``deal facilitating'' (resp. ``deal hindering'') as  they increase (resp. decrease) the TVA and therefore {decrease (resp. increase)} the price (cost of the hedge) for the bank.
Beliefs regarding the tangibility of a benefit-at-own-default{,  which
%in fact
depends in reality on the ability to hedge and therefore monetize this benefit before the actual default,
are} controlled by the choice of the ``own recovery-rate'' parameters $\rho$ and $\mathfrak{r}$.
Larger $\rho$ and $\mathfrak{r}$ {mean smaller} DVA and LVA and therefore {smaller} TVA, {which in principle means less deals} (or {a} recognition of a higher cost of the deal).
This is illustrated numerically in two alternative short rate models to emphasize
the model-free feature of the numerical TVA computations through nonlinear regression BSDE schemes.
The results show that the TVA model risk is under reasonable control {for both} co-calibrated models
(models calibrated to the same initial term structure, but also with the same level of volatility as
imposed through calibration to {cap} prices).
We emphasize however that the latter observation applies to the ``standard'' case studied in this paper without dominant wrong-way and gap risks, two important features which will be dealt with in future work.

\newpage
\bibliographystyle{alpha}
\bibliography{ref}

\newcommand{\etalchar}[1]{$^{#1}$}
\begin{thebibliography}{CAC{\etalchar{+}}10}

\bibitem[BBCH13]{BieleckiBrigoCrepeyHerbertsson13}
T.~R. Bielecki, D.~Brigo, S.~Cr\'epey, and A.~Herbertsson.
\newblock {\em Counterparty Risk Modeling -- Collateralization, Funding and
  Hedging}.
\newblock CRC Press/Taylor \& Francis, 2013.
\newblock In preparation.

\bibitem[BCPP11]{BrigoCapponiPallaviciniPapatheodorou11}
D.~Brigo, A.~Capponi, A.~Pallavicini, and V.~Papatheodorou.
\newblock {Collateral Margining in Arbitrage-Free Counterparty Valuation
  Adjustment including Re-Hypotecation and Netting}.
\newblock Preprint, arXiv:0812.4064, 2011.

\bibitem[BK11]{BurgardKjaer10}
C.~Burgard and M.~Kjaer.
\newblock {PDE Representations of Options with Bilateral Counterparty Risk and
  Funding Costs}.
\newblock {\em The Journal of Credit Risk}, 7(3):1--19, 2011.

\bibitem[BMP13]{BrigoMoriniPallavicini12}
D.~Brigo, M.~Morini, and A.~Pallavicini.
\newblock {\em Counterparty Credit Risk, Collateral and Funding with pricing
  cases for all asset classes}.
\newblock Wiley Finance, 2013.
\newblock Forthcoming.

\bibitem[BP08]{BrigoPallavicini08}
D.~Brigo and A.~Pallavicini.
\newblock {Counterparty Risk and Contingent CDS under correlation between
  interest-rates and default}.
\newblock {\em Risk Magazine}, pages February 84--88, 2008.

\bibitem[CAC{\etalchar{+}}10]{Cesarietal10}
G.~Cesari, J.~Aquilina, N.~Charpillon, Z.~Filipovic, G.~Lee, and I.~Manda.
\newblock {\em Modelling, Pricing, and Hedging Counterparty Credit Exposure}.
\newblock Springer Finance, 2010.

\bibitem[CD12]{CrepeyDouady12}
S.~Cr\'epey and R.~Douady.
\newblock {The Whys of the LOIS: Credit Risk and Refinancing Rate Volatility}.
\newblock SSRN eLibrary, 2012.

\bibitem[CGN12]{CrepeyGrbacNguyen11}
S.~Cr\'epey, Z.~Grbac, and H.~N. Nguyen.
\newblock {A multiple-curve HJM model of interbank risk}.
\newblock {\em Mathematics and Financial Economics}, 6 (3):155--190, 2012.

\bibitem[Cr\12a]{Crepey11b1}
S.~Cr\'epey.
\newblock {Bilateral Counterparty risk under funding constraints -- Part I:
  Pricing.}
\newblock {\em Mathematical Finance}, 2012.
\newblock Forthcoming.

\bibitem[Cr\12b]{Crepey11b2}
S.~Cr\'epey.
\newblock {Bilateral Counterparty risk under funding constraints -- Part II:
  CVA.}
\newblock {\em Mathematical Finance}, 2012.
\newblock Forthcoming.

\bibitem[Cr\12c]{Crepey13}
S.~Cr\'epey.
\newblock Counterparty wrong way and gap risks modeling: A marked default time
  approach.
\newblock In preparation, 2012.

\bibitem[EPQ97]{ElkarouiPengQuenez97}
N.~{El Karoui}, S.~Peng, and M.-C. Quenez.
\newblock {Backward stochastic differential equations in finance}.
\newblock {\em Mathematical Finance}, 7:1--71, 1997.

\bibitem[FST10]{FujiiShimadaTakahashi10b}
M.~Fujii, Y.~Shimada, and A.~Takahashi.
\newblock {Collateral Posting and Choice of Collateral Currency}.
\newblock SSRN eLibrary, 2010.

\bibitem[FT11a]{FilipovicTrolle11}
D.~Filipovi\'c and Anders~B. Trolle.
\newblock {The term structure of interbank risk}.
\newblock SSRN eLibrary, 2011.

\bibitem[FT11b]{FujiiTakahashi11}
M.~Fujii and A.~Takahashi.
\newblock {Derivative Pricing under Asymmetric and Imperfect Collateralization
  and CVA}.
\newblock SSRN eLibrary, 2011.

\bibitem[HL12]{HenryLabordere12}
P.~Henry-Labord\`ere.
\newblock {Counterparty Risk Valuation: A Marked Branching Diffusion Approach}.
\newblock Working paper, 2012.

\bibitem[HTF09]{HastieTibshiraniFriedman09}
T.~Hastie, R~Tibshirani, and J.~Friedman.
\newblock {\em The Elements of Statistical Learning: Data Mining, Inference,
  and Prediction}.
\newblock Springer, 2009.

\bibitem[MR05]{MusielaRutkowski05}
M.~Musiela and M.~Rutkowski.
\newblock {\em Martingale Methods in Financial Modelling}.
\newblock Springer, 2nd edition, 2005.

\bibitem[Pit12]{Piterbarg12}
V.~Piterbarg.
\newblock Cooking with collateral.
\newblock {\em Risk Magazine}, pages July 58--63, 2012.

\bibitem[PR05]{DePriscoRosen05}
B.~De Prisco and D.~Rosen.
\newblock Modelling stochastic counterparty credit exposures for derivatives
  portfolios.
\newblock In M.~Pykhtin, editor, {\em Counterparty Credit Risk Modelling: Risk
  Management, Pricing and Regulation}. RISK Books, London, 2005.

\end{thebibliography}

\end{document}